\documentclass[11pt,a4paper]{article}
\pdfoutput=1
\usepackage{subfigure}
\usepackage{amssymb,amsmath}
\usepackage{graphicx}
\usepackage{color}
\usepackage[colorlinks=true
,urlcolor=blue
,citecolor=blue
,linkcolor=blue
,pagecolor=blue
,linktocpage=true
,pdfproducer=medialab
]{hyperref}
\usepackage[a4paper,width=16.25cm,height=21.75cm]{geometry}

\makeatletter \renewcommand{\@dotsep}{10000} \makeatother

\def\be{\begin{equation}}
\def\ee{\end{equation}}
\def\bea{\begin{eqnarray}}
\def\eea{\end{eqnarray}}
\def\bi{\begin{itemize}}
\def\ei{\end{itemize}}

\def\eslt{\not\!\!{E_T}}
\def\eslt{E_T^{\rm miss}}

\def\to{\rightarrow}

\def\ta{\tilde a}

\def\tb{\tilde b}

\def\td{\tilde d}

\def\tst{\tilde t}
\def\ttau{\tilde \tau}

\def\tg{\tilde g}

\def\tq{\tilde q}

\def\tw{\tilde\chi^{\pm}}
\def\tz{\tilde\chi^0}
\def\alt{\lesssim}
\def\agt{\gtrsim}

\newcommand\prd[3]{{\it Phys.\ Rev.\ }{\bf D #1} (#2) #3}
\newcommand\prl[3]{{\it Phys.\ Rev.\ Lett.\ }{\bf #1} (#2) #3}
\newcommand\npb[3]{{\it Nucl.\ Phys.\ }{\bf B #1} (#2) #3}
\newcommand\plb[3]{{\it Phys.\ Lett.\ }{\bf B #1} (#2) #3}
\newcommand\ptp[3]{{\it Prog.\ Theor.\ Phys.\ }{\bf #1} (#2) #3}
\newcommand\jphg[3]{{\it J. Phys.\ }{\bf G #1} (#2) #3}
\newcommand\app[3]{{\it Astropart.\ Phys.\ }{\bf #1} (#2) #3}
\newcommand\jhep[3]{{\it J. High Energy Phys.\ }{\bf #1} (#2) #3}
\newcommand\epjc[3]{{\it Eur.\ Phys.\ J. }{\bf C #1} (#2) #3}
\begin{document}

\begin{titlepage}
\pagestyle{empty}
\baselineskip=21pt
\rightline{UMN--TH--3029/12, FTPI--MINN--12/03}
\vskip 0.2in
\begin{center}
{\LARGE{\bf Sparticle mass spectra from $SU(5)$ SUSY GUT models
with $b-\tau$ Yukawa coupling unification}}
\end{center}

\begin{center}
{\Large
Howard Baer$^{a,}$\footnote{
Email: baer@nhn.ou.edu},
Ilia Gogoladze $^{b,}$\footnote{
Email: ilia@bartol.udel.edu. On leave of absence from:
Andronikashvili Institute of Physics, GAS, Tbilisi, Georgia.},
Azar Mustafayev $^{c,}$\footnote{
Email: mustafayev@physics.umn.edu},\\
Shabbar Raza $^{b,}$\footnote{
Email: shabbar@udel.edu. On study leave from:
Department of Physics, FUUAST, Islamabad, Pakistan.}
and
Qaisar Shafi $^{b,}$\footnote{
Email: shafi@bartol.udel.edu. }
}

\vskip 0.1in

{\it 
$^a$ Department of Physics and Astronomy,
University of Oklahoma, Norman, OK 73019, USA \\
$^b$ Bartol Research Institute, Department of Physics and Astronomy,
University of Delaware, Newark, DE 19716, USA \\
$^c$ William I.~Fine Theoretical Physics Institute, University of Minnesota,
Minneapolis, MN 55455, USA}

\vskip 0.6cm

\section*{Abstract}
\end{center}
\baselineskip=18pt \noindent

Supersymmetric grand unified models based on the gauge group
$SU(5)$ often require in addition to gauge coupling unification,
the unification of $b$-quark and $\tau$-lepton  Yukawa couplings.
We examine $SU(5)$ SUSY GUT parameter space under the condition of
$b-\tau$ Yukawa coupling unification using 2-loop MSSM RGEs including
full 1-loop threshold effects.
The Yukawa-unified solutions break down into two classes.
Solutions with low $\tan\beta\sim 3-11$ are characterized by
$m_{\tg}\sim 1-4$~TeV and $m_{\tq}\sim 1-5$~TeV. Many of these solutions would be beyond LHC reach,
although they contain a light Higgs scalar with $m_h< 123$~GeV and so may be excluded
should the LHC Higgs hint persist.
The second class of solutions occurs at large $\tan\beta\sim 35-60$, and are a subset of
$t-b-\tau$ unified solutions. Constraining only $b-\tau$ unification to $\sim 5\%$
favors a rather light gluino with $m_{\tg}\sim 0.5-2$~TeV, which should ultimately
be accessible to LHC searches. While our $b-\tau$ unified solutions can be consistent
with a picture of neutralino-only cold dark matter, invoking additional moduli or
Peccei-Quinn superfields can allow for all of our Yukawa-unified solutions to
be consistent with the measured dark matter abundance.

\vspace{1cm}
\vfill
\end{titlepage}

\thispagestyle{empty}
\setcounter{page}{0}
\newpage
\hrule
\bigskip
\tableofcontents
\bigskip\bigskip
\hrule

%

\section{Introduction}
\label{sec:intro}
Grand unified theories (GUTs) based upon the Lie group $SU(5)$
are very compelling in that they unify the disparate gauge groups of the
Standard Model (SM) into a theory based upon just a single gauge group~\cite{su5}.
Furthermore, the $SU(5)$ theory provides a rationale for the seemingly ad-hoc
weak-hypercharge assignments of SM matter fields. A third triumph occurs in that
models based on $SU(5)$ unification unify the $b$-quark and $\tau$-lepton Yukawa
couplings at the unification scale $M_{GUT}\sim 10^{16}$~GeV: then, renormalization
group effects provide roughly the correct values of $m_b$ and $m_{\tau}$ 
at low energy scales~\cite{su5reviews}.

Adding supersymmetry to $SU(5)$ grand unified theories seems essential in order to
stabilize the vast hierarchy separating the weak scale from the GUT scale~\cite{witten}.
The additional SUSY degrees of freedom contained in the Minimal Supersymmetric Standard Model (MSSM) 
alter the RG running of gauge couplings below the scale $M_{GUT}$. 
Indeed, the celebrated unification of the three SM gauge couplings within the MSSM 
is often touted as indirect evidence for weak scale SUSY, and for the possible
existence of a SUSY GUT theory~\cite{gauge}. 

Some drawbacks to the SUSY $SU(5)$ theory include the incorporation of a rather awkward
set of Higgs multiplets which are necessary for appropriate GUT symmetry breaking.
Foremost among these problems is embedding the MSSM Higgs doublets $\hat{H}_u$ and 
$\hat{H}_d$ into a ${\bf 5}$ and ${\bf 5^*}$ respectively of $SU(5)$: 
one must then explain why the color triplets obtain GUT scale masses while the MSSM
multiplets receive weak scale masses: the so-called doublet-triplet splitting problem.
In addition, in SUSY GUT theories based in four spacetime dimensions with spontaneous
GUT symmetry breaking via the Higgs mechanism, protons are expected to decay even in the SUSY
theories with rates which now seem excluded by experiment~\cite{mp,pdecay}.

However, if one formulates $SU(5)$ SUSY GUT models in five or more spacetime dimensions,
then the GUT symmetry can alternatively be broken by compactification of 
the extra dimensions on an appropriate manifold such as an orbifold~\cite{5dguts}. 
This method of symmetry breaking can solve the doublet-triplet splitting problem while suppressing
or even eliminating proton decay. Construction of SUSY GUT models in extra dimensions
can solve many problems endemic to 4-d theories while preserving many of the compelling
features of unified theories. Indeed, in string theory it is possible for SUSY GUT theories
to emerge on four or more dimensions as the low energy (GUT scale) effective theory, 
where the 6-7 additional stringy dimensions must be dispensed with anyway~\cite{raby_review}.

In this paper, we seek to avoid the very model-dependent physics associated with 
the GUT sector and possible extra dimensions, and use instead data and some general
$SU(5)$ SUSY GUT characteristics as a guide to what weak scale physics should look like
at colliders such as the LHC. 
We will assume here that nature is described by an $SU(5)$ SUSY GUT theory
at energy scales $Q\sim M_{GUT}\simeq 2\times 10^{16}$~GeV.
At $M_{GUT}$, the MSSM superfields $\hat{Q}$, $\hat{U}^c$ and $\hat{E}^c$ live 
in the antisymmetric {\bf 10} of $SU(5)$: $\hat{\psi}^{ij}$, while the
$\hat{D}^c$ and $\hat{L}$ superfields live in the ${\bf 5^*}$: $\hat{\phi}_i$. 
(Here, $i$ and $j$ are $SU(5)$ indices running from 1-5.)  
The MSSM Higgs doublet $\hat{H}_d$ is an element of a ${\bf 5^*}$: $\hat{H}_{1i}$, while
$\hat{H}_u$ is an element of a {\bf 5}: $\hat{H}_2^j$. The superpotential $\hat{f}$ then contains
the terms~\cite{pp,bdqt}
\be
\hat{f}\ni \frac{1}{4}f_t\epsilon_{ijklm}\hat{\psi}^{ij}\hat{\psi}^{kl}\hat{H}_2^m +
\sqrt{2}f_b\hat{\psi}^{ij}\hat{\phi}_i\hat{H}_{1j} +\mu_H\hat{H}_{1i}\hat{H}_2^i+ \cdots
\ee 
so that the third generation Yukawa couplings $f_b$ and $f_\tau$ are
unified at $M_{GUT}$, but are distinct from $f_t$.

The soft SUSY breaking terms in an $SU(5)$ SUSY GUT theory are expected to include:
\bea
{\cal L}_{\rm soft}&\ni&-m_{H_1}^2|H_1|^2-m_{H_2}^2|H_2|^2-
m_5^2|\phi |^2 - m_{10}^2tr\{\psi^\dagger\psi \}-
{1\over 2}m_{1/2}\bar{\lambda}_{\alpha} \lambda_{\alpha} \nonumber \\
&+&\left[ {1\over 4}A_tf_t\epsilon_{ijklm}\psi^{ij}\psi^{kl}H_2^m
+\sqrt{2}A_b f_b\psi^{ij}\phi_iH_{1j}+{\it h.c.}\right] .
\eea

For $SU(5)$ SUSY GUT models with $b-\tau$ Yukawa coupling unification, we will adopt a 
GUT scale parameter space given by
\be
m_{5},\ m_{10},\ ,\ m_{H_u}^2,\ m_{H_d}^2,\ m_{1/2},\ A_t,\ A_b,\ \tan\beta, \ sign(\mu )
\label{eq:pspace}
\ee
where we identify $m_{H_u}^2\equiv m_{H_2}^2$ and $m_{H_d}^2\equiv m_{H_1}^2$.
We also take the top quark pole mass to be $m_t=173.3$~GeV, in accord with recent 
measurements from CDF and D0~\cite{mtop}.

Recent previous work on Yukawa coupling unification has focused on $t-b-\tau$ unification 
which is expected to occur in the simplest $SO(10)$ SUSY GUT models~\cite{old}. In these models,
unification of all matter (super)fields of a single generation into a 16-dimensional spinor $\hat{\psi}$
occurs. The two MSSM Higgs multiplets are also unified into a 10-dimensional Higgs representation $\hat{\phi}$.
In these models, it was found that $t-b-\tau$ Yukawa unification can occur if the soft SUSY breaking (SSB) 
parameters are related as $A_0^2\simeq 2m_{10}^2\simeq 4m_{16}^2$~\cite{bf,bdr,abbbft,bkss}. 
With matter scalar SSB masses in the multi-TeV range
and gaugino mass $m_{1/2}$ as small as possible, these relations lead to a weak scale 
sparticle mass spectrum of the inverted mass hierarchy type~\cite{imh}: first/second generation squarks in the $5-20$~TeV
range while third generation scalars are at $\alt 1$~TeV, as required by naturalness. Either a ``just-so''
splitting of Higgs SSB terms (with $m_{H_d}^2>m_{H_u}^2$ at the GUT scale), or $D$-term splitting of scalars (in the DR3
model~\cite{dr3}) is required for radiative electroweak symmetry breaking (REWSB)~\cite{mop}. 

In SUSY models with $t-b-\tau$ Yukawa unification and unified gaugino masses, there is a tendency in the sparticle
mass spectrum for rather light gluinos with $m_{\tg}\alt 500$~GeV~\cite{bf,abbbft,bkss} (although solutions can
also be found with significantly higher gluino masses~\cite{brs}). Recent searches for 
gluino pair production in Yukawa-unified models require $m_{\tg}\agt 500$~GeV, placing some stress 
on this class of models~\cite{atlas}.
One path to relieve such stress is to assume a two-stage breaking pattern 
\be
SO(10)\to SU(5)\to SU(3)_C\times SU(2)_L\times U(1)_Y .
\label{eq:2stage}
\ee
In this case, one might expect a high degree of $b-\tau$ Yukawa unification, but perhaps a lesser degree
of $t-b-\tau$ unification. 
If the scales of the two stages are significantly separated, then one needs to take into account the evolution of
the SSB terms above $M_{GUT}$. Such super-GUT effects can lead to sufficiently different sparticle spectra with
interesting phenomenology~\cite{pp,bdqt,supergut}: for example, the no-scale scenario can be made compatible with
experimental constraints~\cite{noscale}.
The breaking pattern (\ref{eq:2stage}) may also allow for heavier gluino masses to occur in the
range of $m_{\tg}\sim 0.5-1$~TeV. Such gluino masses should be accessible to LHC SUSY searches
with $\sqrt{s}=7$~TeV and $20-30$~fb$^{-1}$ of integrated luminosity~\cite{lhc7}. 

Spurred by these developments, the authors of~\cite{Gogoladze:2009ug,Gogoladze:2010fu} investigated $t-b-\tau$ Yukawa unification  in the framework of
SUSY  $SU(4)_c \times SU(2)_L \times SU(2)_R$~\cite{pati}
(4-2-2, for short). The 4-2-2 structure allows one to have 
non-universal gaugino masses while preserving Yukawa unification. An
important conclusion reached in Ref.~\cite{Gogoladze:2009ug} is that with the same sign but non-universal soft gaugino masses, Yukawa unification in 4-2-2 for $\mu >0$ is compatible with neutralino
dark matter, with gluino co-annihilation~\cite{Gogoladze:2009ug, Profumo:2004wk} playing an important role.

By considering opposite sign gauginos with
{$\mu<0,M_2<0,M_3>0$}, (where $M_2$ and $M_3$ are the SSB gaugino mass terms corresponding  to $SU(2)_L$ and
$SU(3)_c$ respectively, it is shown in Ref.~\cite{Gogoladze:2010fu} that Yukawa
coupling unification consistent with the experimental constraints
can be implemented in 4-2-2. With $\mu<0$ and opposite sign gauginos ($M_2<0$, $M_3>0$),
Yukawa coupling unification is achieved for $m_{16} \gtrsim 300\, {\rm GeV}$, as opposed to 
$m_{16} \gtrsim 8\, {\rm TeV}$ for the case of
same sign gauginos. The finite corrections to the b-quark
mass play an important role here~\cite{Gogoladze:2010fu}. Note that with $M_2 <0$, $M_3>0$ and $\mu<0$,
we can obtain the desired contribution to
$(g-2)_\mu$~\cite{Bennett:2006fi}. This enables us to simultaneously
satisfy the requirements of  $t-b-\tau$ Yukawa unification in 4-2-2,
neutralino dark matter and $(g-2)_\mu$, as well as a variety of
several other bounds.

Encouraged by the abundance of solutions and co-annihilation channels
available in the case of Yukawa unified SUSY 4-2-2, Yukawa unification in SO(10) GUT was explored in~\cite{Gogoladze:2011ce} with non-universal MSSM gaugino masses at $ M_{\rm GUT}$. This scenario can arise from non-singlet F-terms, compatible with the underlying GUT symmetry~\cite{Gogoladze:2011ce}. Furthermore, the soft masses for the two scalar Higgs doublets are set equal ($m_{H_{u}}$=$m_{H_{d}}$) at $M_{\rm GUT}$. It is intriguing to note that in these models, rather precise $t-b-\tau$ Yukawa unification also happens to
yield a mass for the lightest CP-even Higgs boson in the  $122 -124$~GeV range~\cite{bf,Gogoladze:2011aa}.
There is an approximately 2~GeV theoretical uncertainty in this calculation.

The remainder of this paper is organized as follows. In Sec.~\ref{sec:calc}, we outline details of
our sparticle mass spectra calculation, along with the requirement of $b-\tau$ Yukawa unification,
and constraints from earlier collider and $B$ decay searches. In Sec.~\ref{sec:results}, we show
preferred $SU(5)$ model parameter choices which lead to $b-\tau$ Yukawa unification. The solutions divide
into two classes: 1. those with low $\tan\beta\sim 3-11$ for which the top Yukawa coupling 
$f_t\gg f_b\simeq f_\tau$ at the GUT scale, and 2. those at high $\tan\beta\sim 35-60$, which give 
$t-b-\tau$ Yukawa quasi-unification, {\it i.e.} $f_t\sim f_b\simeq f_\tau$ at $Q=M_{GUT}$. 
The low $\tan\beta$ solutions may be eliminated if the LHC hint of Higgs at $m_h\simeq 125$ holds true.
Otherwise, LHC direct searches for sparticles can only cover a portion of the low $\tan\beta$ solutions, 
while searches for gluino pair production at LHC can cover nearly all of the high $\tan\beta$ solutions.
In Sec.~\ref{sec:dm}, we discuss aspects of the relic abundance of dark matter for $b-\tau$ Yukawa-unified models.
While neutralino-only dark matter can be accommodated by a variety of co-annihilation, resonance annihilation or
higgsino annihilation processes, generically we expect a standard overabundance of neutralinos. 
Either an overabundance or an under abundance of neutralino dark matter can be brought into accord with the
measured CDM abundance by invoking either additional late-decaying scalar (moduli) fields, or by invoking
a Peccei-Quinn axion superfield (containing axion, saxion and axino components) which is needed anyway 
as a solution to the strong $CP$ problem.
A summary and conclusions are presented in Sec.~\ref{sec:conclude}.


\section{Calculation of sparticle mass spectra with $b-\tau$ Yukawa unification
\label{sec:calc}}

For our calculations, we adopt the Isajet 7.80~\cite{isajet,bfkp} SUSY spectrum generator Isasugra.
Isasugra begins the calculation of the sparticle mass spectrum with
input $\overline{DR}$ gauge couplings and $f_b$, $f_\tau$ Yukawa couplings at the 
scale $Q=M_Z$ ($f_t$ running begins at $Q=m_t$) and evolves the 6 couplings up in energy 
to scale $Q=M_{GUT}$ (defined as the value $Q$ where $g_1=g_2$) using two-loop RGEs. 
We do not strictly enforce the unification condition $g_3 = g_1 = g_2$ at $M_{GUT}$, 
since a few percent deviation from unification can be assigned to unknown GUT-scale 
threshold corrections~\cite{Hisano:1992jj}.
At $Q=M_{GUT}$, the SSB boundary conditions are input, and  the set
of 26 coupled two-loop MSSM RGEs~\cite{mv} are evolved back down in scale to
$Q=M_Z$. 
Full two-loop MSSM RGEs are used for soft term evolution, while the gauge and Yukawa coupling
evolution includes threshold effects in the one-loop beta-functions, so the gauge and
Yukawa couplings transition smoothly from the MSSM to SM effective theories as 
different mass thresholds are passed.
In Isajet 7.80, the values of SSB terms which mix are frozen out at the
scale $Q\equiv M_{SUSY}=\sqrt{m_{\tst_L} m_{\tst_R}}$, while non-mixing SSB terms are frozen out
at their own mass scale~\cite{bfkp}. 
The scalar potential is minimized using the RG-improved one-loop MSSM effective
potential evaluated at an optimized scale $Q=M_{SUSY}$ which accounts for
leading two-loop effects~\cite{haber}.
Once the tree-level sparticle mass spectrum is computed, full one-loop
radiative corrections are calculated for all sparticle and Higgs boson masses,
including complete one-loop weak scale threshold corrections for the
top, bottom and tau masses at scale $Q=M_{SUSY}$~\cite{pbmz}. These fermion self-energy 
terms are critical to evaluating whether or not Yukawa couplings do indeed unify~\cite{hrs}.
Since the GUT scale Yukawa couplings are modified by the threshold corrections, the
Isajet RGE solution must be imposed iteratively with successive up-down 
running until a convergent sparticle mass solution is found.
For most of parameter space, there is excellent agreement between Isajet and
the SoftSUSY, SuSpect and Spheno codes,\footnote{These three codes invoke an ``all-at-once'' transition from
MSSM to SM effective theories in contrast to the Isasugra approach.} although at the edges of parameter
space agreement between the four codes typically diminishes~\cite{kraml}.

We searched for Yukawa-unified solutions in the $SU(5)$ parameter space (\ref{eq:pspace}) in two stages. First,
we performed the MCMC scan~\cite{bkss,mcmc} over the large parameter range
\bea
m_{10},\ m_5,\ m_{H_u},\ m_{H_d}:\ 0-20\ {\rm TeV},\\
m_{1/2}:  0-2\ {\rm TeV},\\
-60\ {\rm TeV}<A_t,\ A_b<60\ {\rm TeV},\\
\tan\beta :\ 1.1-60 .
\label{eq:prange}
\eea
We identify several solutions with good $b-\tau$ Yukawa unification ($R\leq 10\%$) and the neutralino 
Relic Density $\Omega_{\tz_1}h^2$ within the
 WMAP bound~\cite{wmap7}. Those solutions become centers of second-stage scans with narrower parameter ranges,
where we look for more solutions with good $b-\tau$ Yukawa unification, good $\Omega_{\tz_1}h^2$ and try to make the spectra as light as possible.
Each generated parameter set is entered into Isasugra using the 
non-universal SUGRA model inputs, and our initial selection criteria is that the points
generate a neutralino $\tz_1$ as lightest MSSM particle, and appropriate REWSB.
In plots to follow, these points are labeled with gray color. 

We next require the following bounds (inspired by LEP2/Tevatron searches) on sparticle masses:
\bi
\item $m_h>114.4$~GeV,
\item $m_{\tst_1},\ m_{\tb_1}>100$~GeV,
\item $m_{\ttau_1}>105$~GeV,
\item $m_{\tg}>250$~GeV,
\item $m_{\tw_1}>103$~GeV.
\ei
Using Isatools~\cite{bmm,bsg} and~\cite{mamoudi}, 
we also require the following bounds from heavy flavor ($B$-physics) to be respected:
\bi
\item $BF(B_s\to\mu^+\mu^- )<1.1\times 10^{-8}$ \cite{bmmexp}, 
\item $2.85\times 10^{-4}<BF(b\to s\gamma )< 4.24\times 10^{-4}$ \cite{hfag},
\item $0.15\le BF^{SUSY}(B_u\to\tau\nu_\tau )/BF^{SM}(B_u\to\tau\nu_\tau )\le 2.41$ \cite{hfag}.
\ei
Points passing both mass and $B$-physics cuts are labeled as red or green.

For each solution, we calculate the degree of $b-\tau$ Yukawa unification at the GUT scale via 
the $R$-parameter:
\be
R=\frac{max(f_b,f_\tau )}{min(f_b,f_\tau )} .
\label{eq:R}
\ee
Thus, $R=1.0$ would tag a solution with perfect $b-\tau$ Yukawa coupling unification.

\section{Results
\label{sec:results}}

\subsection{$SU(5)$ parameters required by $b-\tau$ unification
\label{ssec: params}}

\begin{figure}
\centering
\subfiguretopcaptrue
\subfigure{
\includegraphics[width=7.cm]{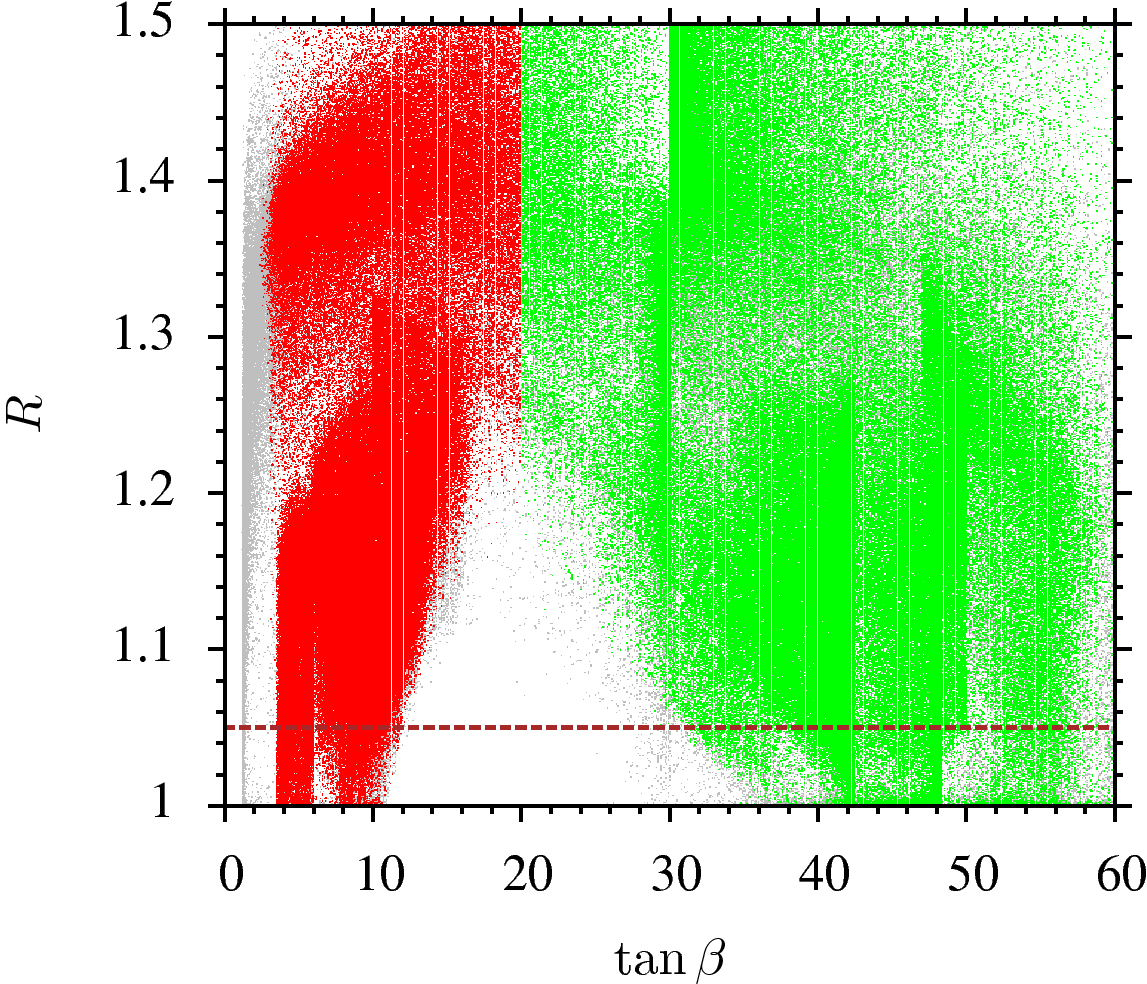}
}
\subfigure{
\includegraphics[width=7.cm]{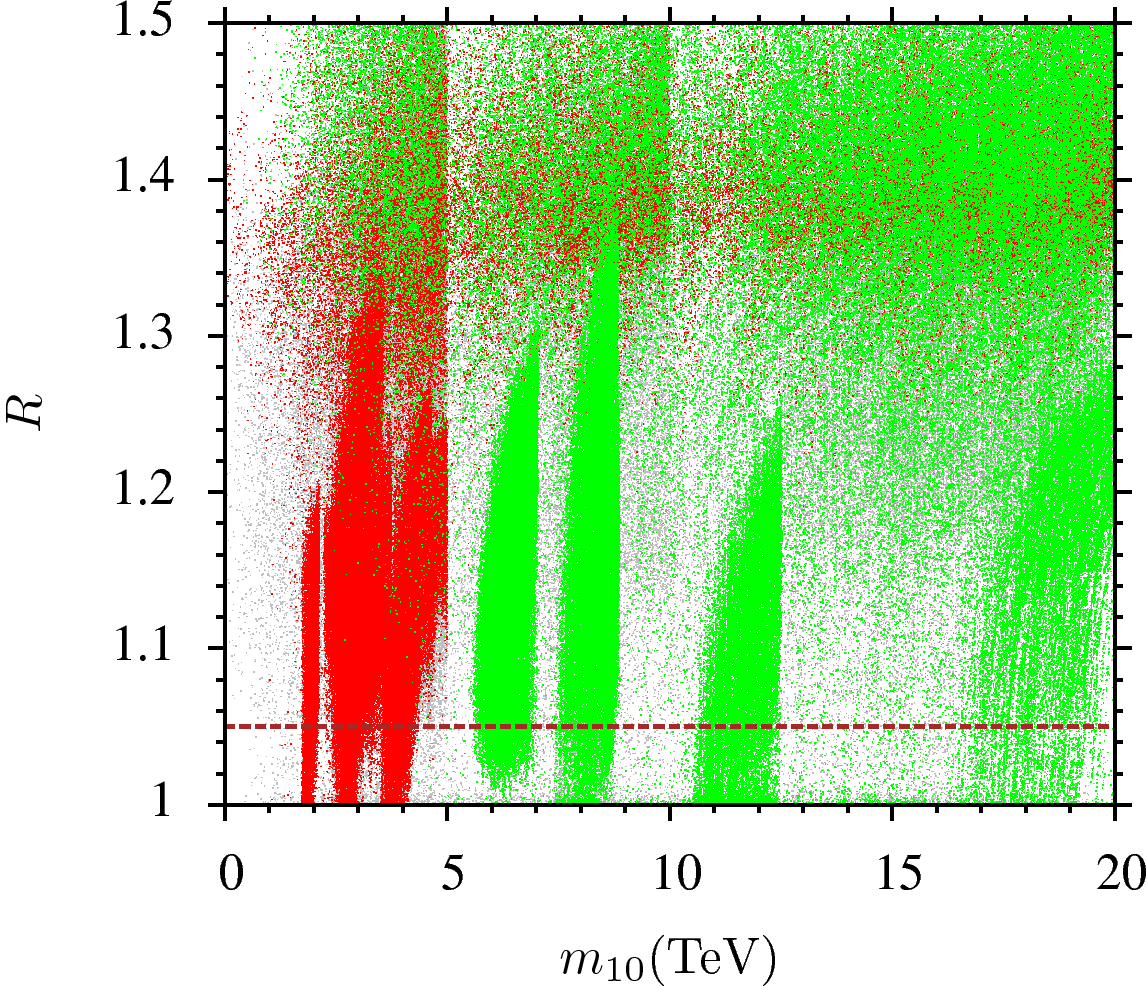}
}
\subfigure{
\includegraphics[width=7.cm]{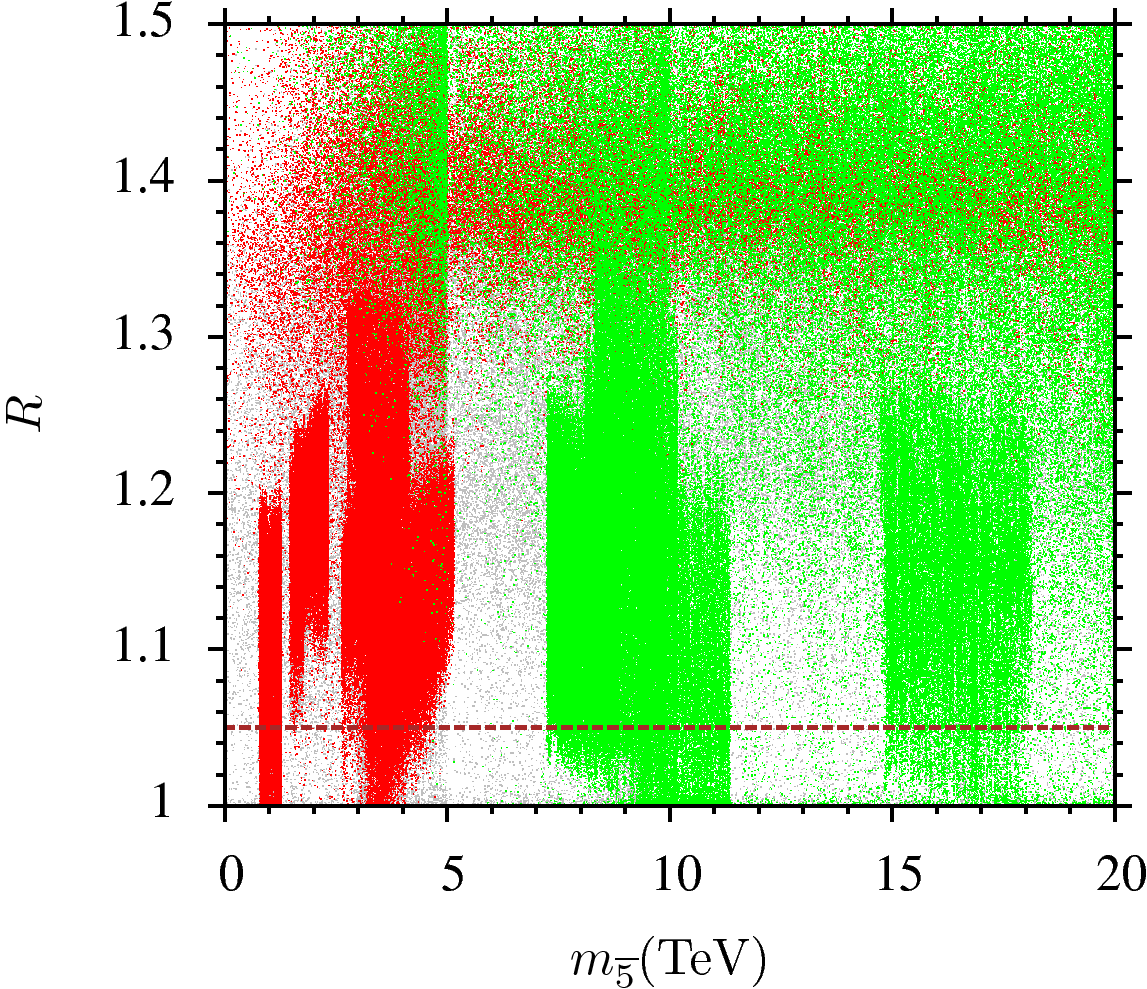}
}
\subfigure{
\includegraphics[,width=7.cm]{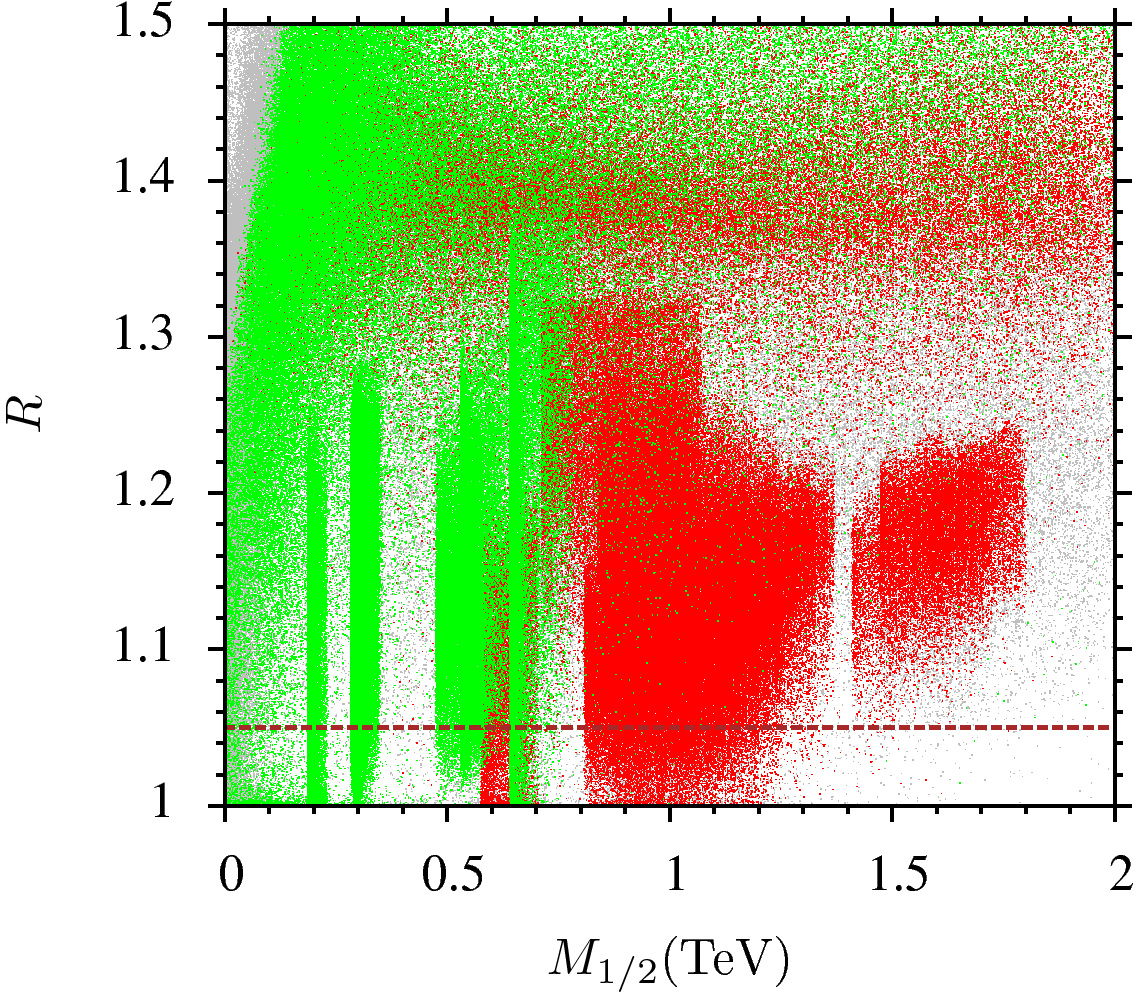}
}
\subfigure{
\includegraphics[totalheight=5.5cm,width=7.cm]{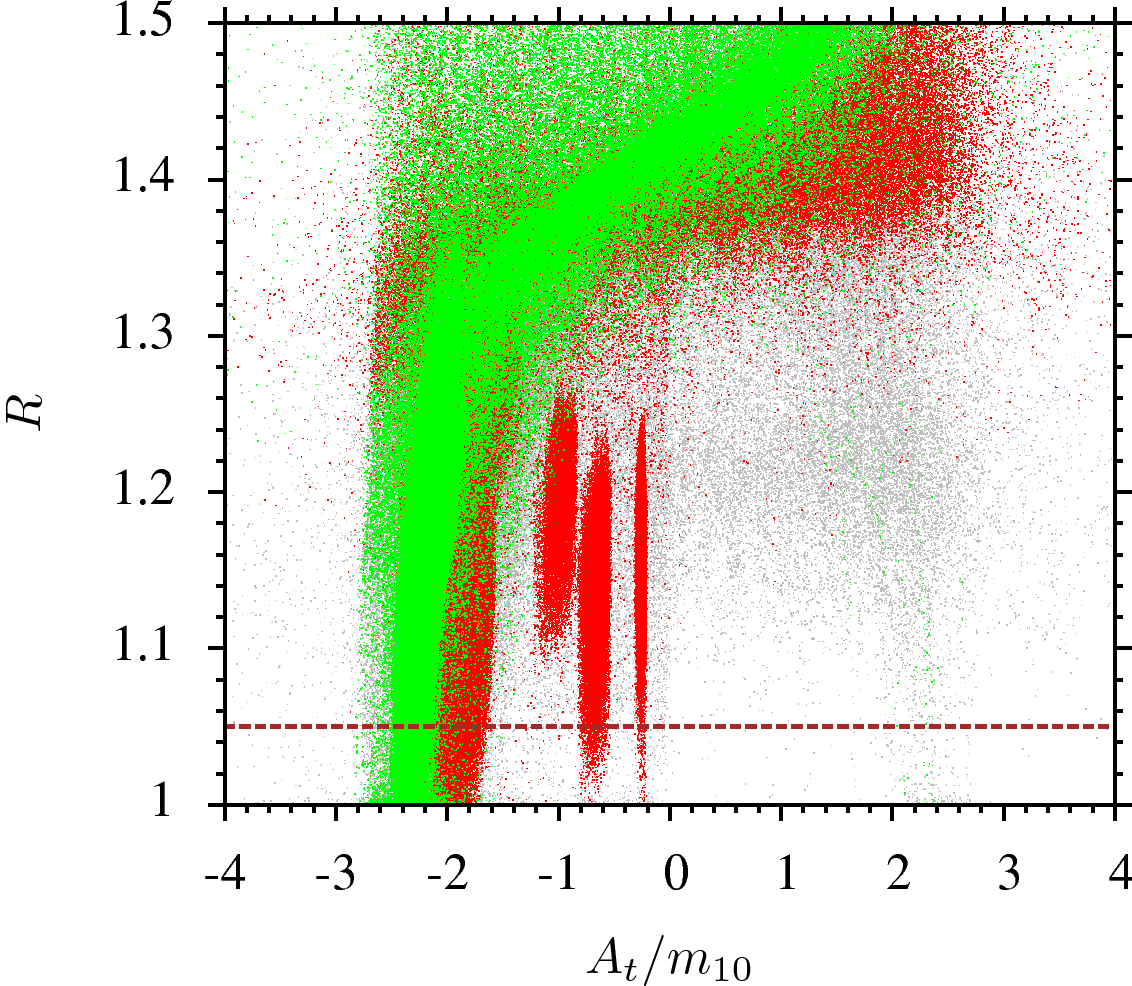}
}
\subfigure{
\includegraphics[totalheight=5.5cm,width=7.cm]{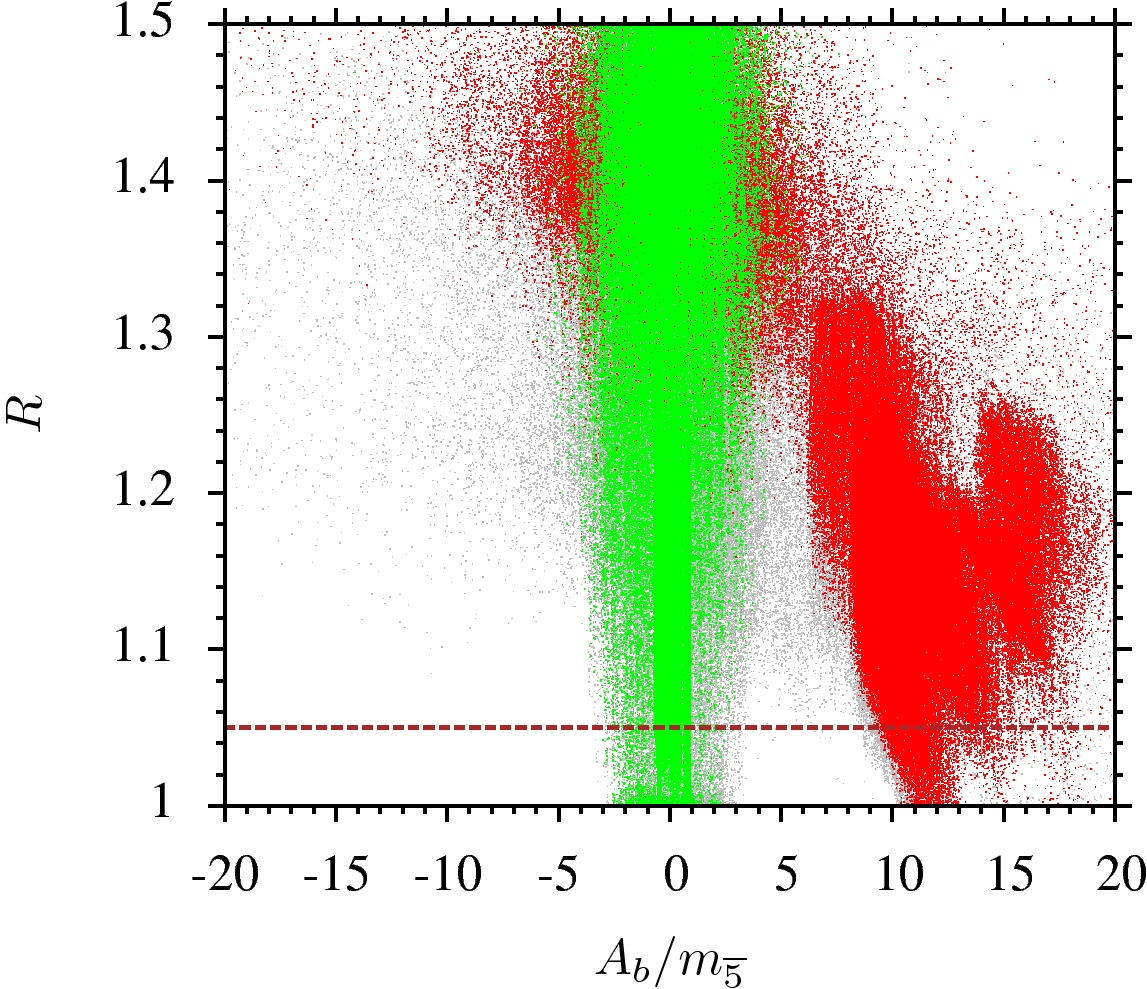}
}
\caption{$b-\tau$ Yukawa unificaion as function of model parameters for $\mu >0$ and $m_t =173.3$~GeV.
Gray points satisfy REWSB and neutralino as LSP conditions.
Red and green points satisfy additional spartcile mass and $B$-physics bounds and have
$\tan\beta <20$ and $\tan\beta >20$, respectively. The horizontal dashed line indicates the 5\% Yukawa
unification. 
}
\label{fig:params}
\end{figure}

Our first results are shown in Fig.~\ref{fig:params}.
In Fig.~\ref{fig:params}{\it a}), we plot the value of $R\ vs.\ \tan\beta$. Points with
$R\alt 1.05$ have a high degree of $f_b-f_\tau$ Yukawa unification. 
We see immediately that the solutions break up into two classes: low $\tan\beta\sim 3-11$ 
and high $\tan\beta\sim 35-60$. We color code the resulting points according to this criteria:
red points have $\tan\beta <20$ and green points have $\tan\beta > 20$.
Apparently no $b-\tau$ unified points can be generated for $11<\tan\beta <35$ which satisfy the
mass and $B$-physics bounds.

We will first discuss the low $\tan\beta\sim 3-11$ solutions. 
From Fig.~\ref{fig:params}{\it b}) and {\it c}), we see that this class of solutions
requires $m_{10}\sim 1-4$~TeV and $m_5\sim 1-5$~TeV. In addition, from frame {\it d}), we see
that rather large values of $m_{1/2}\sim 0.6-1.5$~TeV are required. Such high $m_{1/2}$ values 
lead to gluinos with $m_{\tg}\agt 1.3$~TeV. Moreover, as noted above, if we insist that $R\le 1.05$,   there is a clear distinction between low and high $\tan\beta$ solutions. However, if we relax this condition ($R\agt 1.3$), this distinction between the low and high $\tan\beta$ solutions disappears. 
Another point to note is that here we have $b-\tau$ unified solutions for low $\tan\beta$. Such solutions are ruled out in the CMSSM because of the Higgs mass bound~\cite{Gogoladze:2011be}. In our scans we find solutions with $\tan\beta$ as low as ${\rm 3.5}$. 
This limit can be slightly changed by varying $m_{t}$ and $m_{b}$. From frames {\it e}) and {\it f}), 
we see that $R<1.05$ solutions require the GUT-scale trilinear SSB parameters to be in the range
$-2.3m_{10}<A_t<0$ and $A_b\sim (10-20)m_5$. The latter parameter $A_b(M_{GUT})$ again strongly differentiates
between the low and high $\tan\beta$ solutions. 
The low $\tan\beta$ solutions are characterized by $f_b\simeq f_{\tau}\sim 0.04$ at the GUT scale,
while $f_t(M_{GUT})\sim 0.5$, {\it i.e.} there is a large disparity between $f_t$ and
$f_b\simeq f_\tau$. This class of models might be indicative of an $SU(5)$ GUT theory, but with
no connection to $SO(10)$ (unless we proceed to $SO(10)$ models where the MSSM Higgs doublets
live in separate {\bf 10}s of $SO(10)$~\cite{bdqt}).

In contrast, from Fig.~\ref{fig:params} we see the green points with $\tan\beta \sim 35-60$ 
require $m_{10}\sim m_5\sim 5-20$~TeV, {\it i.e.} these solutions require multi-TeV
matter scalars just as do SUSY models with $t-b-\tau$ Yukawa unification~\cite{abbbft,bkss,alt}.
Furthermore, we see from frame {\it d}) that the high $\tan\beta$ points require much lower values of
$m_{1/2}\alt 0.7$~TeV, which leads to an upper bound on the gluino mass of $m_{\tg}\alt 2$~TeV. 
From frame {\it e}), we find that $-2.8m_{10}<A_t<-1.8 m_{10}$, 
with also a few solutions around $A_t\sim 2.2 m_{10}$.
In frame {\it f}), we see that $A_b$ is much less correlated, with $|A_b|\alt 3m_5$.
The high $\tan\beta\sim 45-55$ solutions also tend to have a high degree of $f_t\simeq f_b\simeq f_\tau\sim 0.55$
unification, whereas the solutions with $\tan\beta\sim 35-45$ tend to have  $f_t\sim 0.55$, but
$f_b\simeq f_\tau\sim 0.47$. These latter solutions might be indicative of an $SO(10)$ SUSY GUT which 
has broken to $SU(5)$ at a higher mass scale than where $SU(5)$ breaks to the SM gauge group.

In Table~\ref{tab:bm}, we list low $\tan\beta$ and high $\tan\beta$ benchmark solutions for illustration. 
Points 1 and 2 belong to the set of low $\tan\beta$ Yukawa unified points and also represent
$A$-resonance and sbottom co-annihilation scenarios. Points 3 and 4 are representative of high $\tan\beta$ Yukawa unified solutions, where point 3 depicts a stop co-annihilation solution, while point 4 shows a large
$\Omega h^2$ value. 

\begin{table}[h!]
\centering
\begin{tabular}{lcccc}
\hline
\hline
                  & Point 1 & Point 2 & Point 3 & Point 4  \\
\hline
$m_{10}$           & 2604  & 3849 & 18380  & 16800 \\
$m_{\overline{5}}$ & 3443  & 900.1  & 16450 & 18960  \\
$m_{1/2}$          & 1049  & 1056 & 292.6  &  358.6\\
$\tan\beta$       &  8.3   & 4.77  & 42.4 &  45\\
$A_{t}$           & -5140  & -7455 & -4484 &   -39510\\
$A_{b}=A_{\tau}$   & 41070  & 40830 & -8170  &   23640 \\
$m_{H_d}$          & 3424  &  905   & 1850  & 17340 \\
$m_{H_u}$          & 1380  &  4700   & 14150  & 10410 \\
$sign(\mu)$       & +      & +  & + & + \\
\hline
$f_t(M_{GUT})$     & 0.496  & 0.518  & 0.555 & 0.567 \\
$f_b(M_{GUT})$     & 0.058  & 0.033  & 0.474 & 0.542 \\
$f_{\tau}(M_{GUT})$& 0.059  & 0.034  & 0.485 & 0.542\\
\hline
$m_h$             & 120.9  & 119.6  & 125.1 & 125.2    \\
$m_{A}$           & 929  & 797  & 18781 &  13544 \\
$\mu$             & 2934  & 2345  & 17562 &  17394 \\
\hline
$m_{\tilde{\chi}^0_{1,2}}$ &461, 882   &467, 887 &179, 362 &179, 354   \\
$m_{\tilde{\chi}^0_{3,4}}$ &2857, 2859  &2291, 2295 &16905, 16905 &16406, 16406    \\
$m_{\tilde{\chi}^{\pm}_{1,2}}$ &881, 2857  &887, 2311 &368, 17075 &357, 16429    \\
$m_{\tilde{g}}$    & 2385  & 2431  & 1089  & 1165 \\
\hline 
$m_{ \tilde{u}_{L,R}}$  &3314, 3211   &4336, 4405 & 18374, 18265 & 16788, 16608    \\
$m_{\tilde{t}_{1,2}}$   &1211, 1798   &1007, 2825 & 215, 10165 & 3289, 7153   \\
\hline 
$m_{ \tilde{d}_{L,R}}$ &3315, 3984  &4337, 2033  &18374, 16488 & 16788, 19095    \\
$m_{\tilde{b}_{1,2}}$  &1375, 2082  &489, 2841 &10198, 11734 &7139, 12709      \\
\hline
$m_{ \tilde{e}_{L,R}}$  &3479, 2719  &1321, 3731 &16319, 18556 & 18850, 17052    \\
$m_{\tilde{\tau}_{1,2}}$ &876, 2939  &803, 341 &14263, 14864 & 11256, 16464   \\
\hline
$\Omega_{\tz}h^2$  &0.113   & 0.074 & 0.11 & 2269  \\
$\langle\sigma v\rangle(v\rightarrow 0)\ [cm^3/s]$ & 3.886$\times 10^{-27}$ & 9.512$\times 10^{-29}$ & 1.684$\times 10^{-26}$ 
                          & 4.385$\times 10^{-31}$  \\
$\sigma^{SI}(\tz p)\times 10^{12}$ [pb]  & 5.639 & 9.689 & 1.640 & 0.127 \\
\hline
$a_\mu^{SUSY} \times 10^{10}$  & 0.134   & 0.239 & 0.015 & 0.013  \\
$BF(b\rightarrow s\gamma )\times 10^4$  & 3.319 & 3.501 & 3.059 & 3.038  \\
$BF(B_S\rightarrow \mu\mu )\times 10^9$  & 3.826 & 3.838 & 3.867 & 3.903  \\
\hline
\hline
\end{tabular}
\caption{Input parameters and resulting mass spectra  and rates for several sample points from the scan. 
All masses and dimensionful parameters are in GeV units. 
}
\label{tab:bm}
\end{table}

\subsection{$SU(5)$ preferred masses
\label{ssec: mass}}

\begin{figure}
\centering
\subfiguretopcaptrue
\subfigure{
\includegraphics[totalheight=5.5cm,width=7.cm]{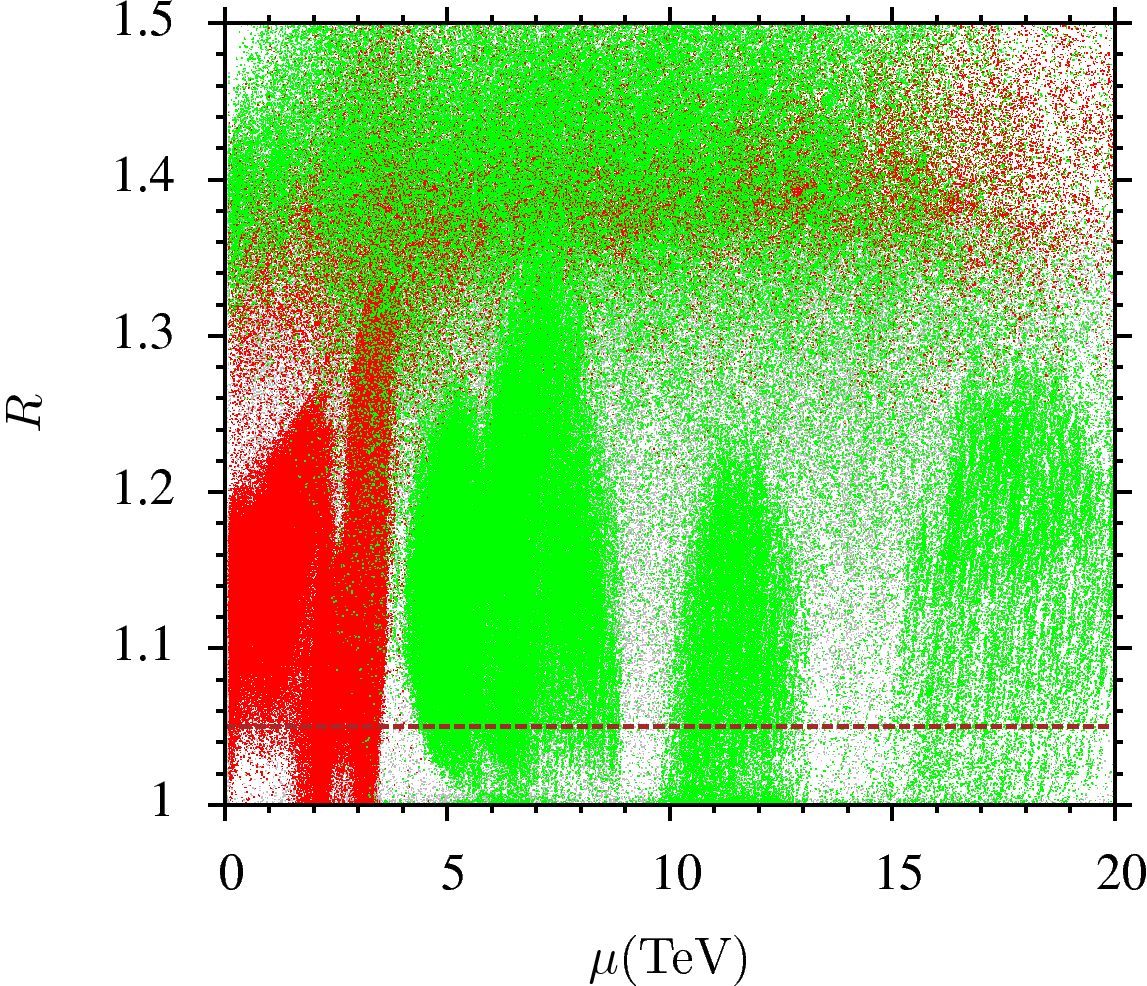}
}
\subfigure{
\includegraphics[totalheight=5.5cm,width=7.cm]{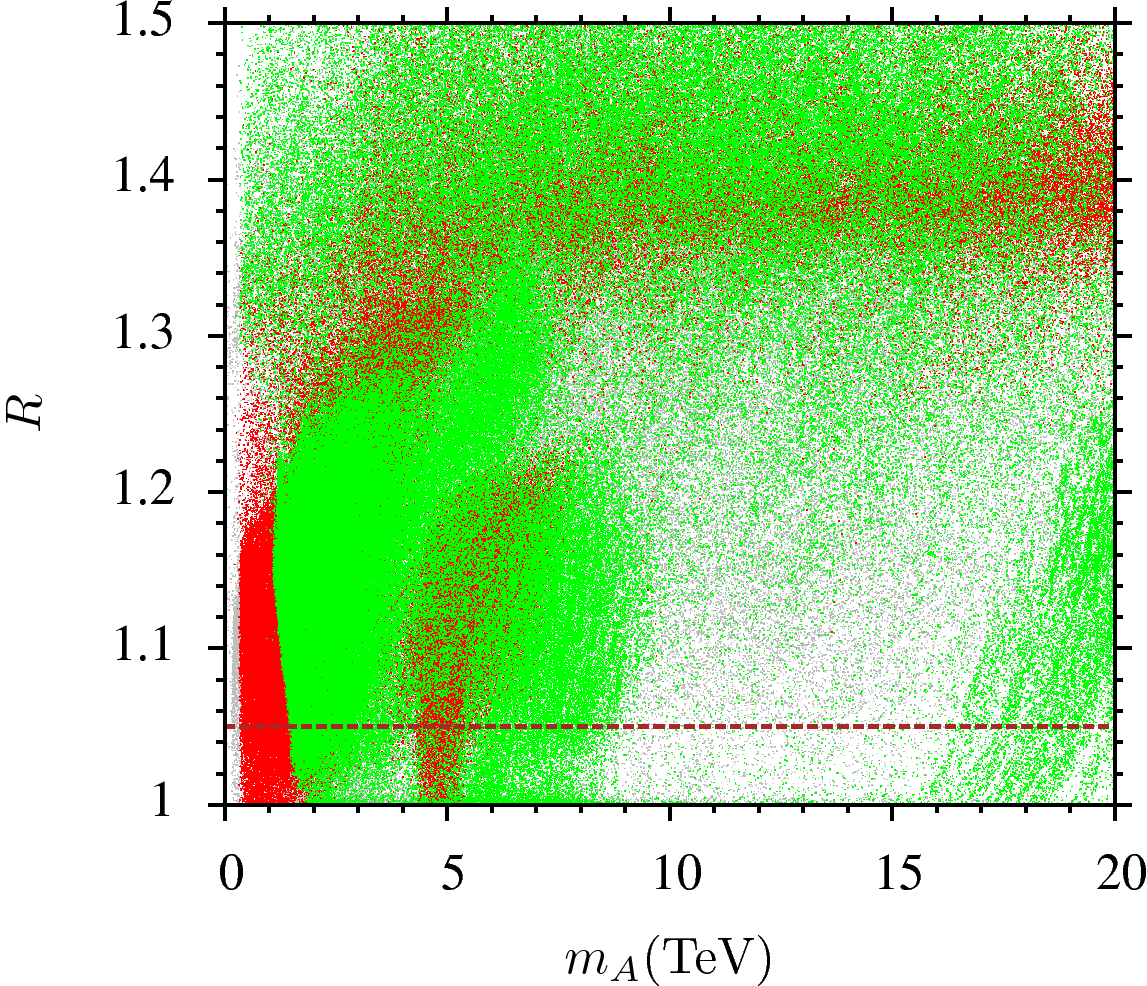}
}
\subfigure{
\includegraphics[totalheight=5.5cm,width=7.cm]{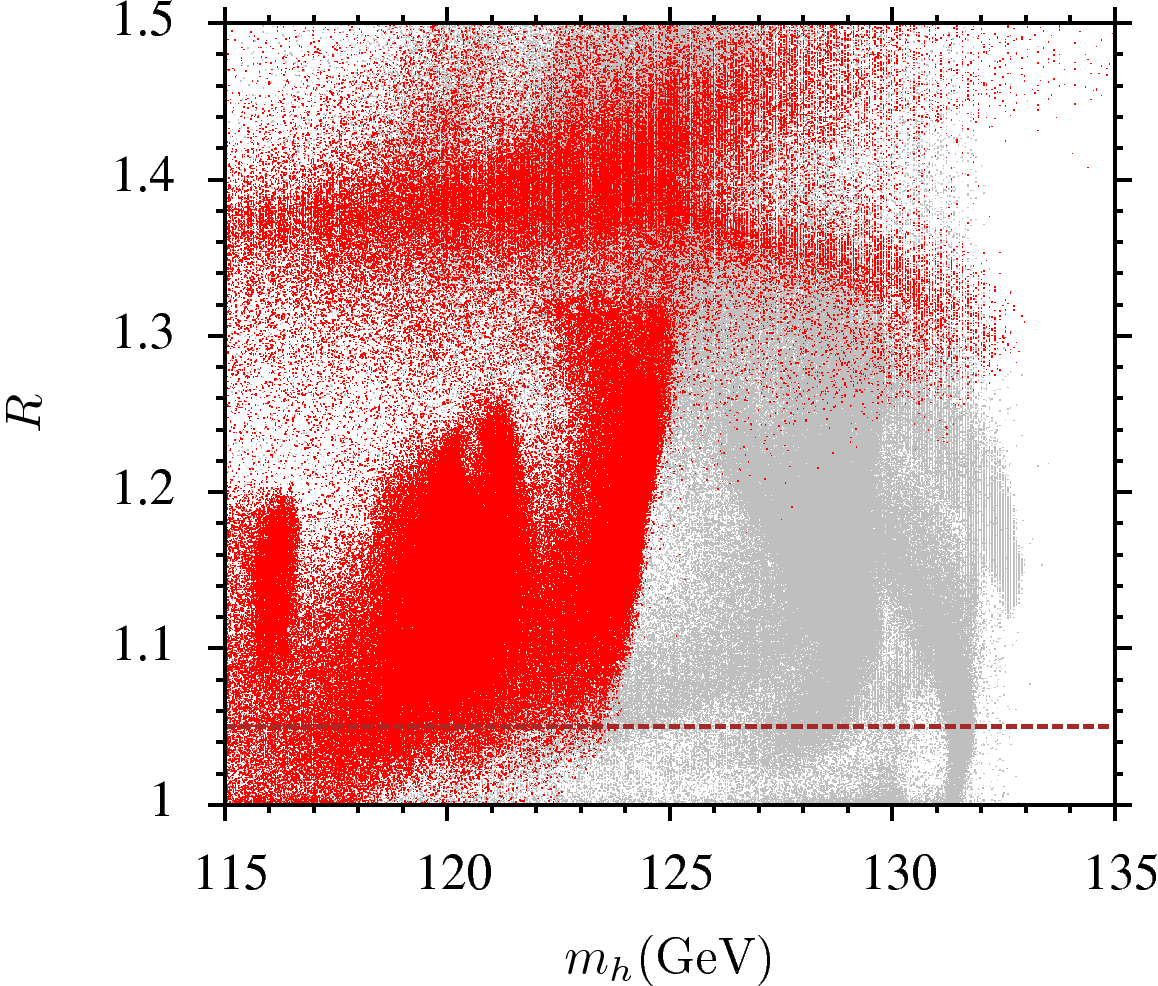}
}
\subfigure{
\includegraphics[totalheight=5.5cm,width=7.cm]{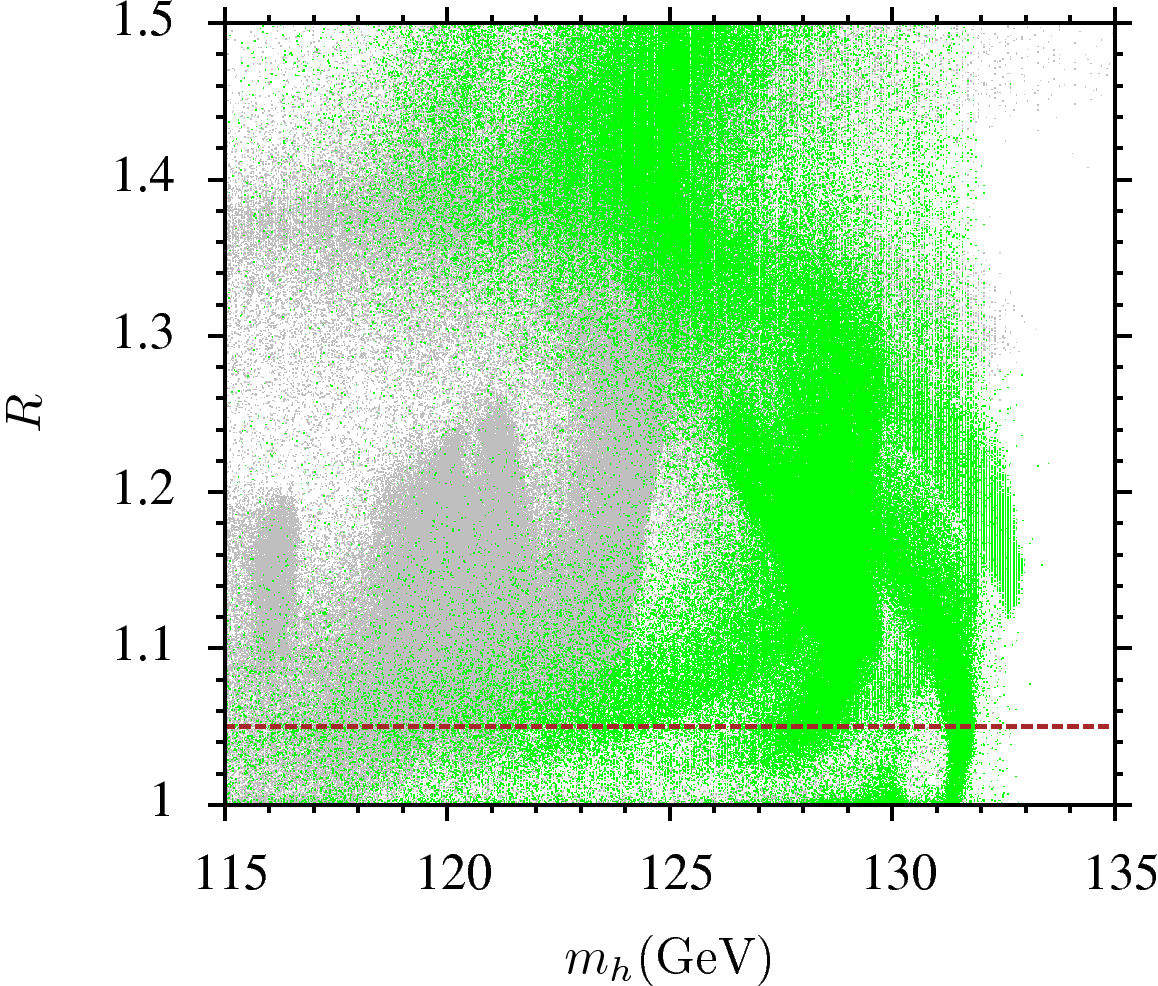}
}
\caption{$R_{b\tau}$ versus parameters in the Higgs sector for $\mu >0$ and $m_t =173.3$~GeV.
The color coding is the same as in Fig.~\ref{fig:params}.
}
\label{fig:mass}
\end{figure}

We next proceed to examine some derived parameters associated with the Higgs/higgsino sector
from SUSY models with $b-\tau$ unification.
In Fig.~\ref{fig:mass}{\it a}), we show the correlation of the degree of the Yukawa unification with the parameter $\mu$. 
The magnitude of $\mu$ is determined by minimization conditions on the Higgs scalar potential.
Here, we see that the low $\tan\beta$ solutions also give rise to a range $\mu\sim 0-5$~TeV:
{\it i.e.} $\mu$ is bounded from above, and furthermore $\mu$ can be well below the 1~TeV scale.
This may allow for the lightest neutralino $\tz_1$ to be of mixed bino-higgsino type, which
gives rise to WMAP-allowed values of thermal neutralino relic abundance. However, the high $\tan\beta$ solutions
require rather large values of $\mu$, typically in the multi-TeV range, so that for these solutions we
would expect the $\tz_1$ state to be a nearly pure bino. 
Here too we can see the separation of low and high $\tan\beta$ solutions for $R\alt 1.05$. 
%

In frame {\it b}), we show correlations of the $R$-parameter with the mass of the {\it CP}-odd Higgs boson $A$.
The low $\tan\beta$ solutions
require $m_A\sim 0-6$~TeV. The rather low range of $m_A$ may allow for neutralino
annihilation through the $A$-resonance in the early universe. Note that for low $\tan\beta$ values, we can have LHC accessible solutions for $m_A$ if we require $b-\tau$ unification better than 5\%. 
Meanwhile, the high $\tan\beta$ solutions tend to have $m_A$ inhabiting the multi-TeV range, 
so that $A$ resonance annihilation is unlikely to be a possibility.

In frames {\it c}) and {\it d}), we plot $R$ as function of the mass of the light $CP$-even Higgs boson $h$  
for low $\tan\beta$ (red) and high $\tan\beta$ (green) solutions, respectively. 
In this case, we see that the low $\tan\beta$
solutions require $m_h\alt 123$~GeV, with $m_h$ usually much lower. Meanwhile, the large 
$\tan\beta$ solutions allow for $m_h\sim 123-133$~GeV. 
Recently, some evidence has been reported from the Atlas and CMS experiments~\cite{atlas_h,cms_h,h125} 
for a SM-like Higgs boson very near to $\sim 125$~GeV. If this result is maintained by the
factor of 4-6 more data from LHC expected in 2012, then it would likely {\it rule out} the
low $\tan\beta$ $b-\tau$ unified solutions, while maintaining consistency with the high $\tan\beta$
solutions!

Because good $b-\tau$ Yukawa unification requires large values of SSB masses, $m_5 \gtrsim 1$~TeV, first and
second generation sleptons are rather heavy with masses greater than $\sim 1$~TeV ($\gtrsim 4$~TeV for high
$\tan\beta$ values). This leads to the large suppression of the SUSY contribution to the muon anomalous
magnetic moment $(g-2)_\mu$ that arizes at 1-loop level from diagrams involing smuon and muon sneutrino. As
result $a_\mu^{SUSY}$ values, that we computed using the IsaAMU~\cite{gm2_th} subroutine, is always several orders of
magnitude below the extracted discrepancy $\Delta a_\mu = (28.7\pm 8.0)\times 10^{-10}$~\cite{davier}. This can be seen in
several sample points we listed in Table~\ref{tab:bm}.

\subsection{Prospects for LHC SUSY searches
\label{ssec: lhc}}

\begin{figure}
\centering
\subfiguretopcaptrue
\subfigure{
\includegraphics[width=10.0cm]{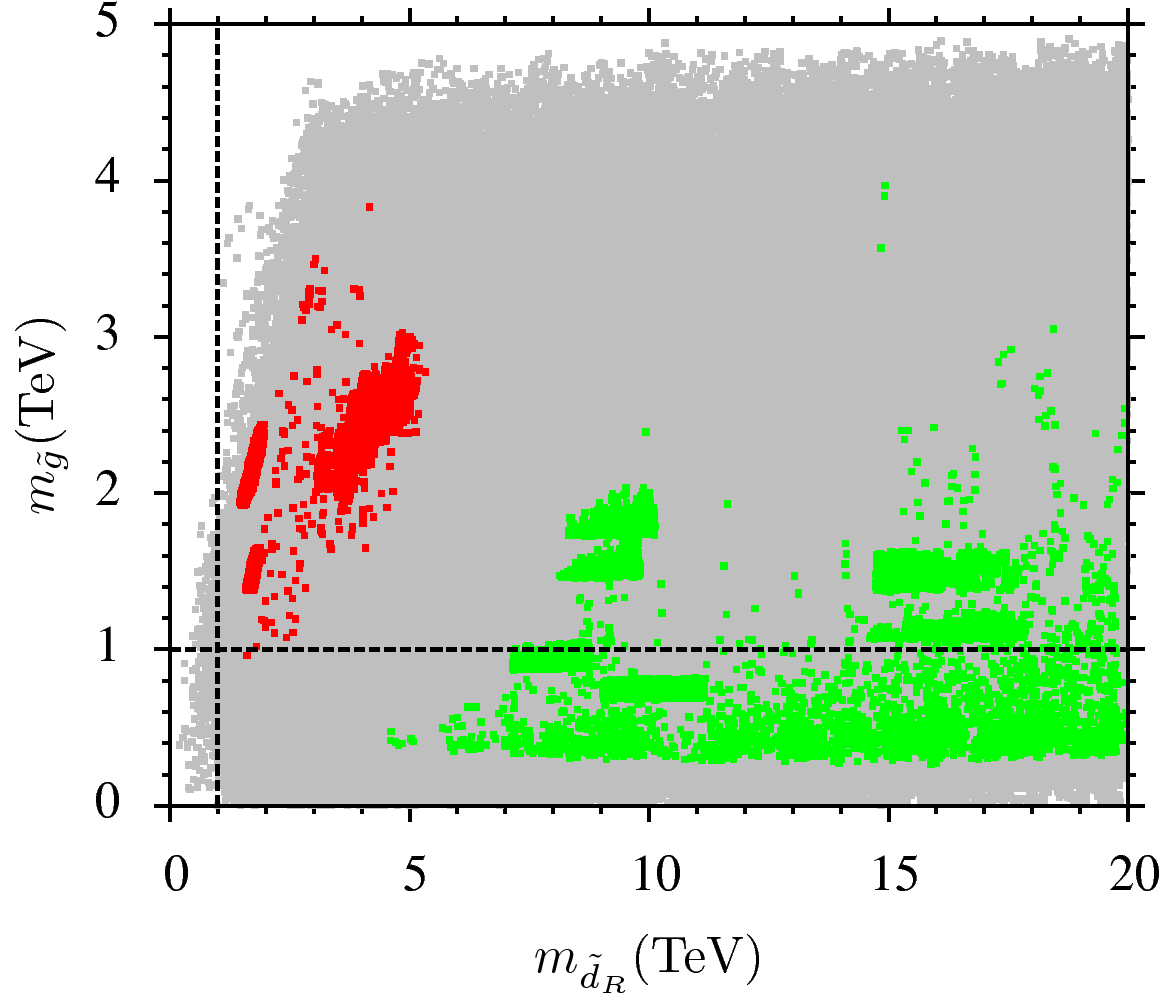}
}
\caption{Distribution of points from the scan in the mass plane of gluino and 1st/2nd generation
squarks for $\mu >0$ and $m_t =173.3$~GeV.
Gray points satisfy REWSB and neutralino as LSP conditions.
Red and green points satisfy additional mass bounds, $B$-physics bounds, have $R<1.05$ and represent
$\tan\beta <20$ and $\tan\beta >20$, respectively.
The dashed lines represent approximate reaches for LHC7.
}
\label{fig:lhc}
\end{figure}

In this section, we discuss prospects for detection of $b-\tau$ Yukawa-unified SUSY
at the CERN LHC $pp$ collider with either $\sqrt{s}=7$ or 14~TeV. 
In Fig.~\ref{fig:lhc}, we show solutions which pass the mass and $B$-physics cuts -- but also with
$R<1.05$ -- in the $m_{\td_R}\ vs.\ m_{\tg}$ plane. 
The value of $m_{\td_R}$ is meant to exhibit a typical first/second generation squark mass.
The red low $\tan\beta$ points all have $m_{\tg}>1$~TeV with $m_{\tq}\sim 1.5-5$~TeV. 
In this case, we find $m_{\tq}\sim m_{\tg}$ or slightly heavier. The reach of LHC with $\sqrt{s}=7$~TeV
and 20~fb$^{-1}$ (LHC7) extends to $m_{\tg}\sim m_{\tq}\sim 1.5$~TeV~\cite{lhc7}, 
while LHC with $\sqrt{s}=14$~TeV and 100~fb$^{-1}$ extends to $m_{\tg}\sim m_{\tq}\sim 3$~TeV~\cite{lhc14}. 
Thus, about half the red points will be within reach of LHC14, 
while those with $m_{\tq}>3$~TeV and $m_{\tg}\agt 2$~TeV will likely be beyond LHC14 reach.
LHC luminosity or energy upgrades will be necessary to probe more deeply into the low $\tan\beta$
Yukawa-unified space of solutions.

In the case of high $\tan\beta$ solutions (green points), we see that $m_{\tq}$ exceeds -- 
and frequently far exceeds -- 5~TeV. Meanwhile, the gluino mass is bounded from above, with $m_{\tg}$
almost always $<2$~TeV. 
In this case, LHC searches will focus on gluino pair production~\cite{so10lhc}. 
For $m_{\tq}\gg m_{\tg}$, the LHC7 reach is to $m_{\tg}\sim 1$~TeV~\cite{lhc7}, 
while LHC14 reach is to $m_{\tg}\sim 1.7$~TeV~\cite{lhc14}. 
Recent work on LHC signatures in the case where $m_{\tq}\gg m_{\tg}$ implies the maximal LHC14 reach
in models with gaugino mass unification is to $m_{\tg}\sim 2$~TeV in the 
$\tw_1\tz_2\to Wh+\eslt$ channel~\cite{wh}.
Thus, we expect LHC7 to probe the region $m_{\tg}\alt 1$~TeV, and LHC14 to nearly cover the 
remaining parameter space with 100-1000~fb$^{-1}$ of integrated luminosity.

\section{Dark matter relic density in $b-\tau$ unified models
\label{sec:dm}}

\begin{figure}
\centering
\subfiguretopcaptrue
\subfigure{
\includegraphics[width=10.0cm]{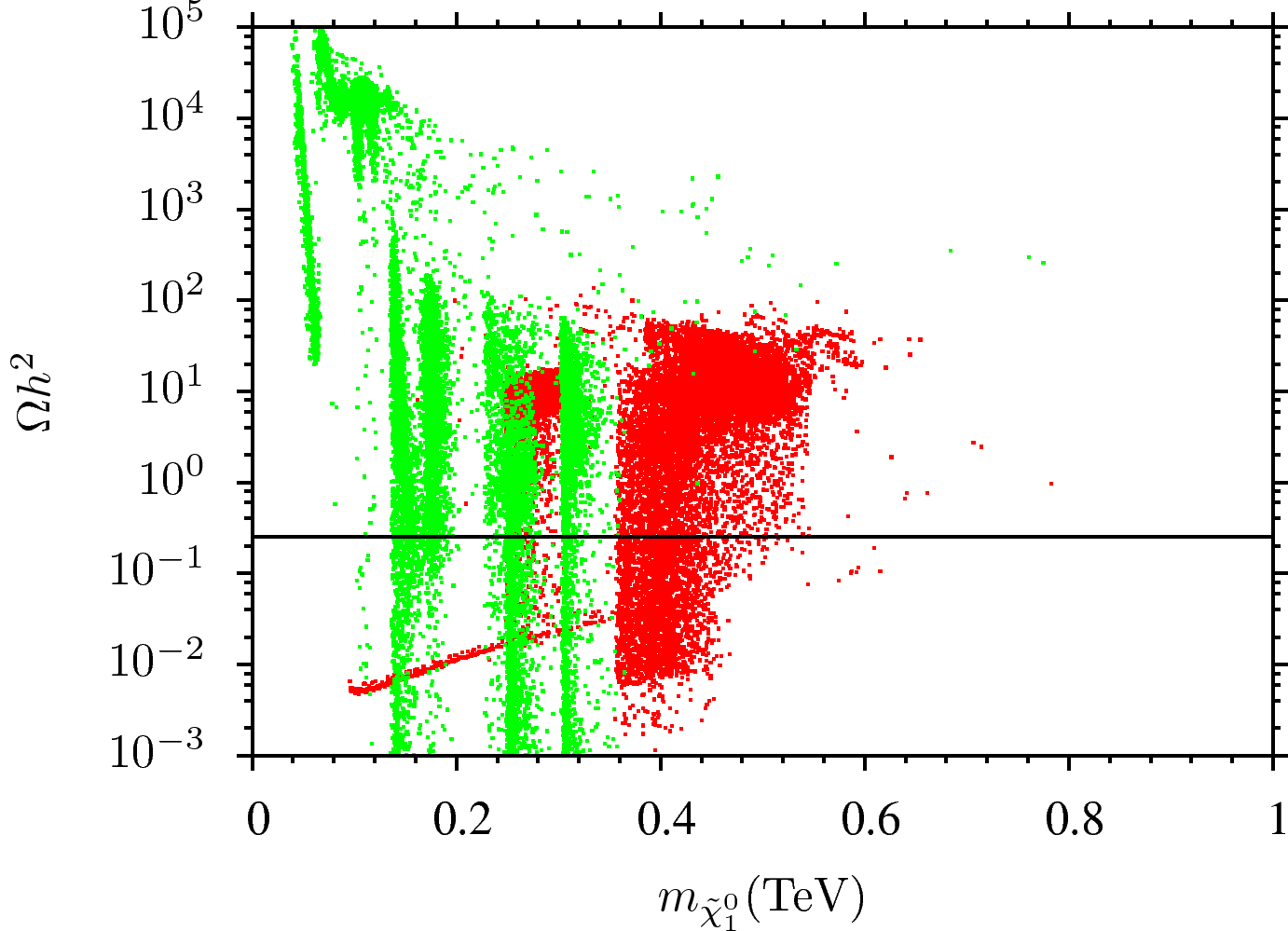}
}
\caption{Neutralino relic density $\Omega_{\tz_1}h^2$ versus the neutralino mass $m_{\tz_1}$ from the scan
with $\mu >0$.
All points satisfy mass bounds, $B$-physics bounds and have $R<1.05$. 
Red and green color represent solutions with $\tan\beta <20$ and $\tan\beta >20$, respectively.
The solid horizontal line represents the WMAP measured value~\cite{wmap7}.
}
\label{fig:Oh2}
\end{figure}

It has been noted long ago that $t-b-\tau$ Yukawa-unified models tended to give
a huge overabundance of thermally produced neutralino-only dark matter~\cite{abbbft}.\footnote{
In Ref.~\cite{roszkowski}, it is suggested that $A$-resonance annihilation may be available
for bring neutralino-only CDM into its measured range. However, a number of other authors have failed
to reproduce $t-b-\tau$ unified solutions with very low $m_A$ values such that $2m_{\tz_1}\sim
m_A$~\cite{abbbft,bkss,codes}.}
To this end, we adopt the IsaReD relic density calculator~\cite{isared} to compute 
the thermally produced neutralino abundance $\Omega_{\tz_1}h^2$ from $b-\tau$ unified models. 
The value of $\Omega_{\tz_1}h^2$ versus $m_{\tz_1}$ is plotted for solutions with $R<1.05$ 
in Fig.~\ref{fig:Oh2}. The red solutions with low $\tan\beta$
span a range $10^{-3}<\Omega_{\tz_1}h^2<10^2$. The large $\tan\beta$ solutions populate a much larger
range of  $10^{-3}<\Omega_{\tz_1}h^2<10^5$ owing to the suppression of $\tz_1$ pair annihilation by
exchange of multi-TeV scalars in the relevant Feynman diagrams.
This is to be compared with the CDM abundance $\Omega_{CDM}h^2=0.1109\pm 0.0056$ reported by 
WMAP7~\cite{wmap7}, which we indicate as a horizontal black line. 

\begin{figure}
\centering
\subfiguretopcaptrue
\subfigure{
\includegraphics[totalheight=5.5cm,width=7.cm]{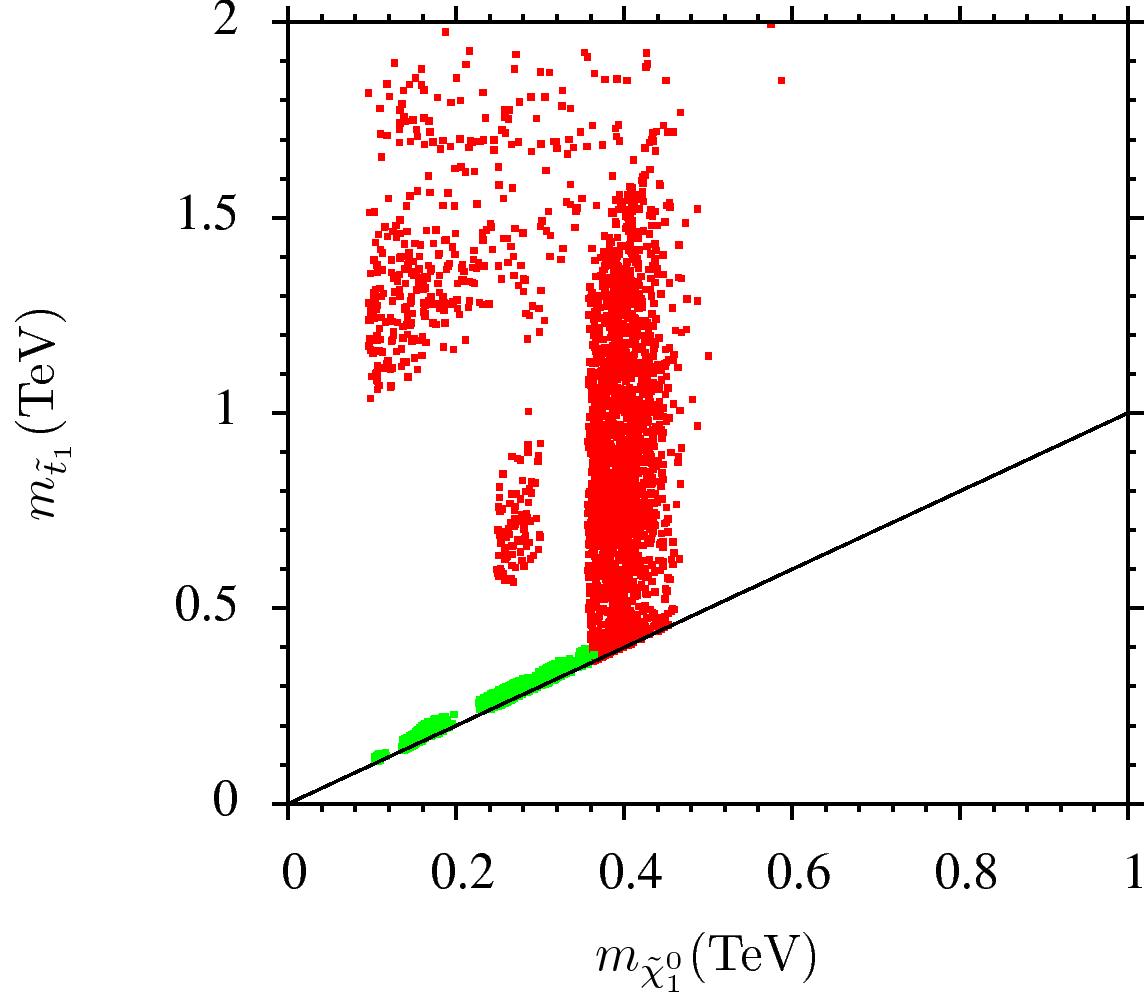}
}
\subfigure{
\includegraphics[totalheight=5.5cm,width=7.cm]{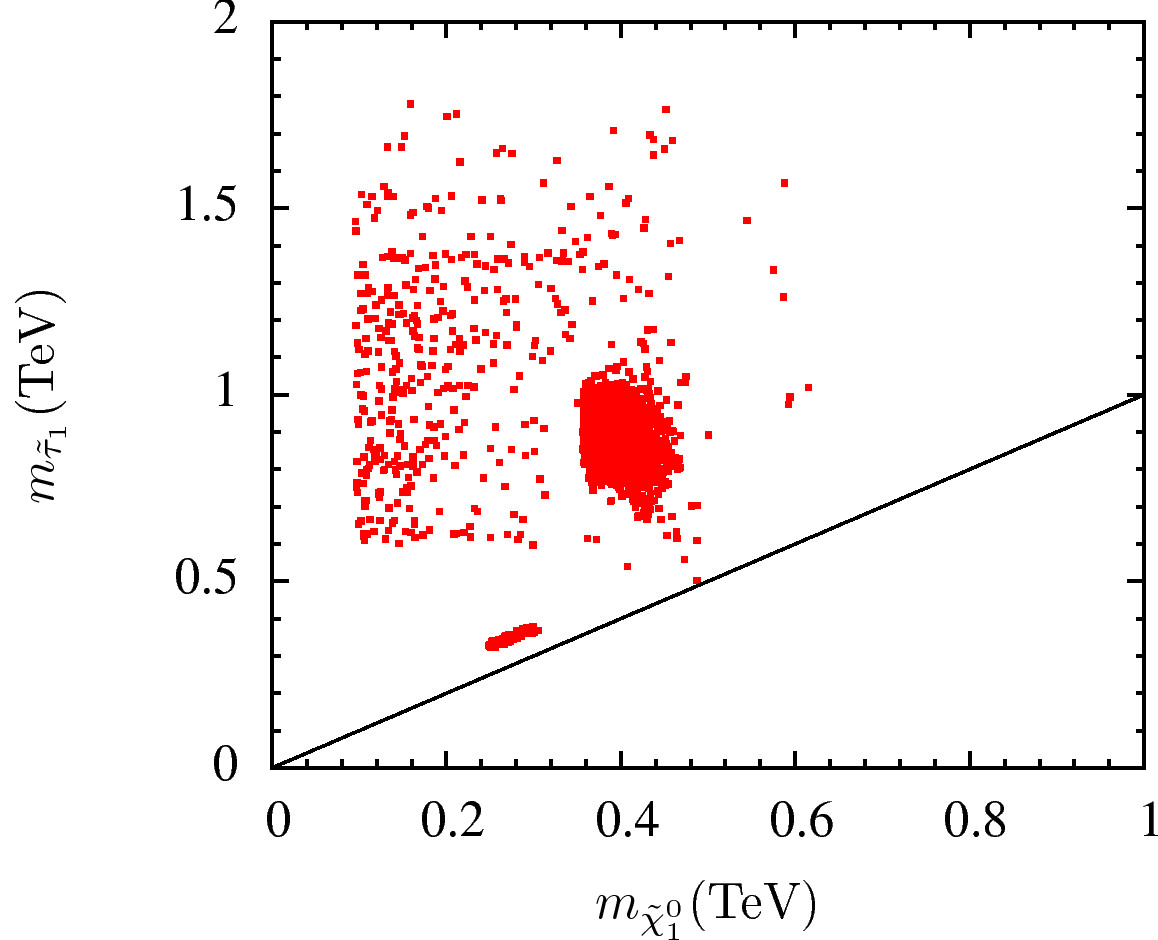}
}
\subfigure{
\includegraphics[totalheight=5.5cm,width=7.cm]{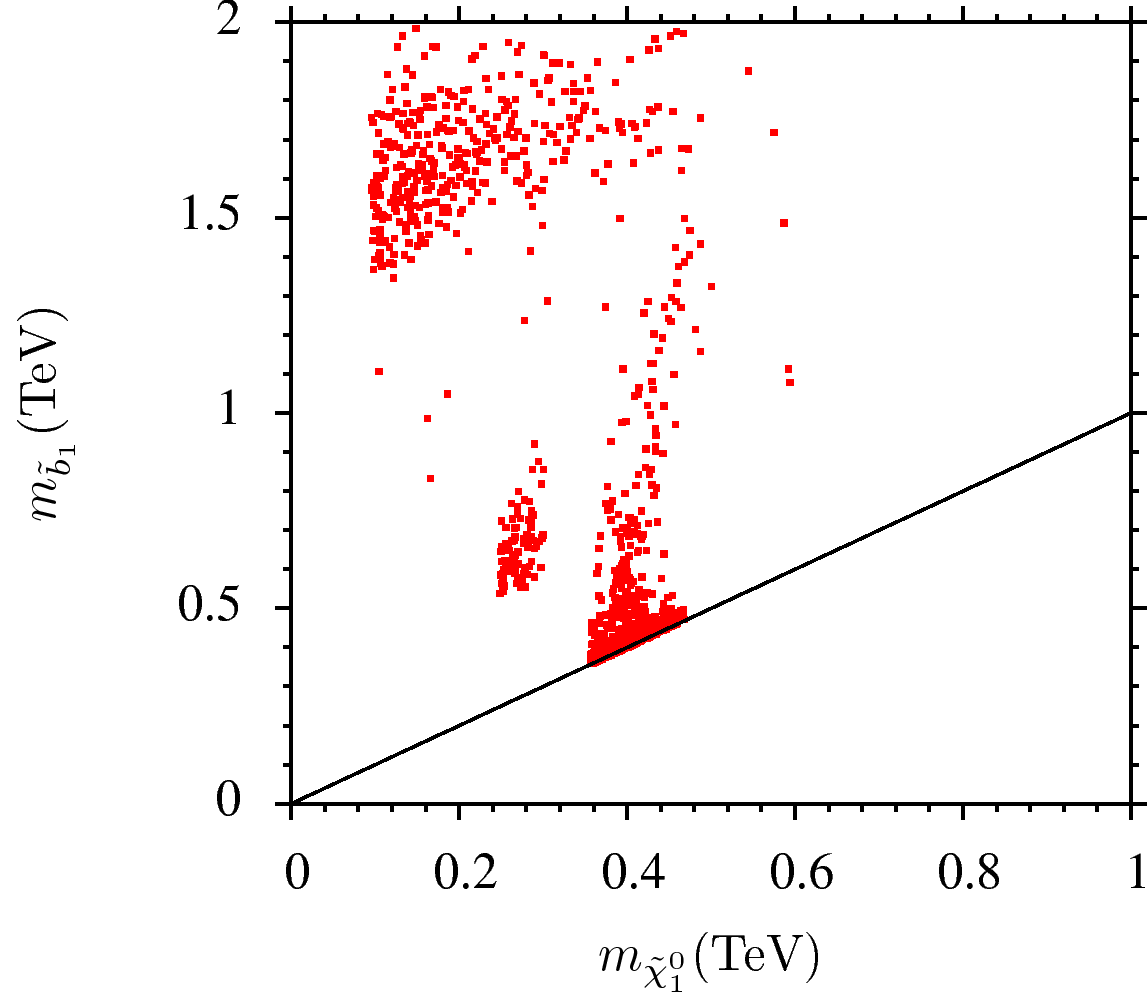}
}
\subfigure{
\includegraphics[totalheight=5.5cm,width=7.cm]{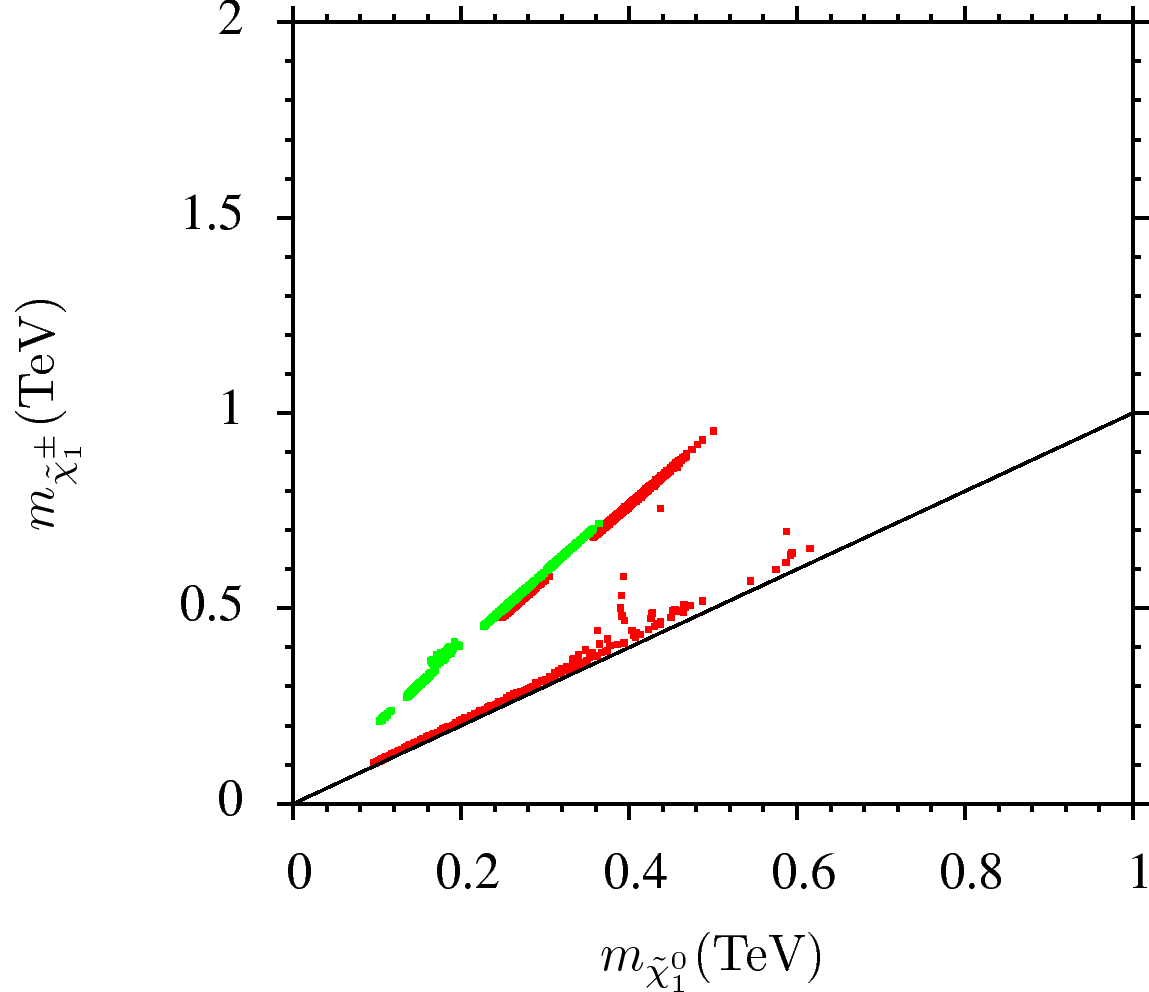}
}
\subfigure{
\includegraphics[totalheight=5.5cm,width=7.cm]{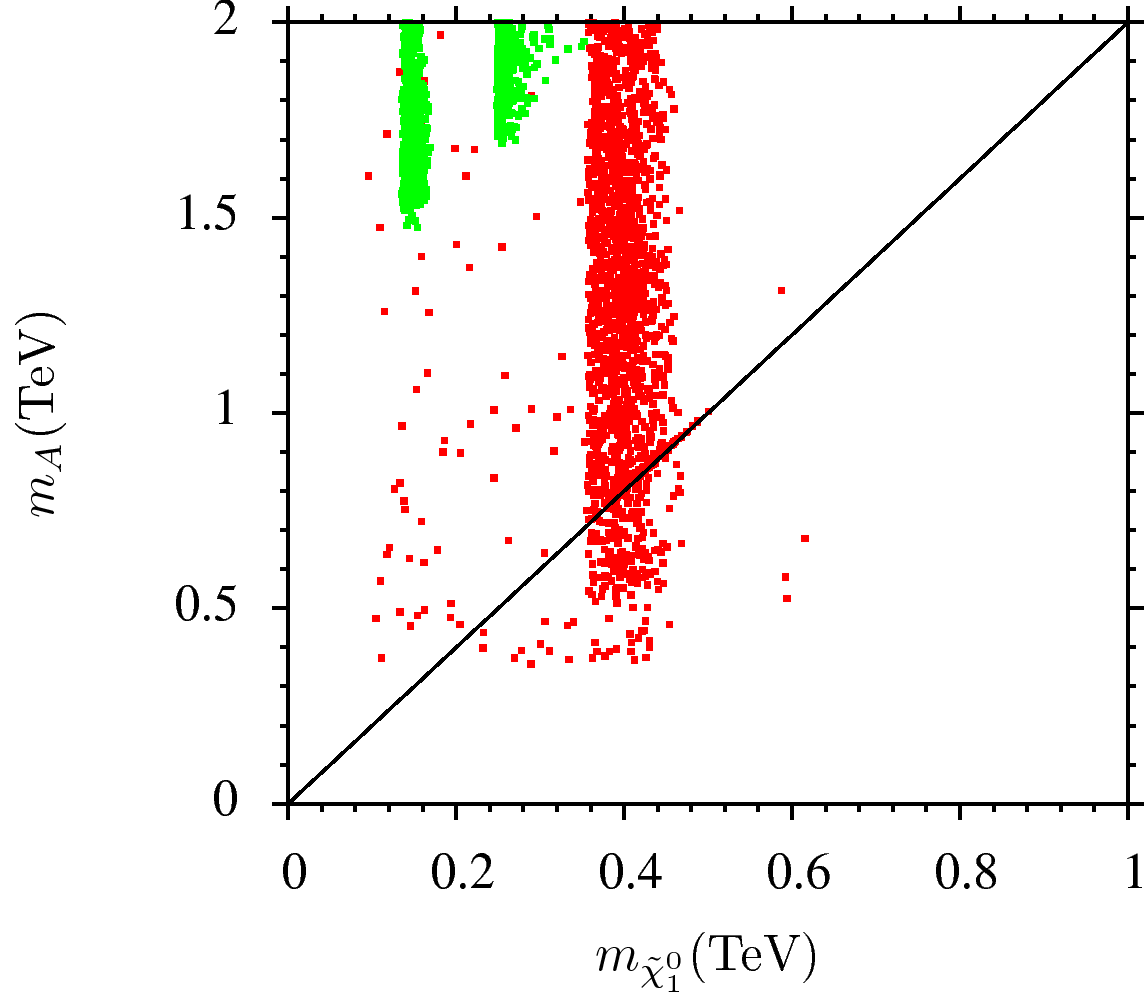}
}
\caption{Distribution of points from the scan in various planes of sparticles/Higgs masses
for $\mu >0$ and $m_t =173.3$~GeV.
All points satisfy mass bounds, $B$-physics bounds and have $R<1.05$ and $\Omega h^2<0.139$.
Red and green color represent solutions with $\tan\beta <20$ and $\tan\beta >20$, respectively.
Diagonal black lines represent approximate mass conditions for sizable coannihilation or resonance pair
annihilation of neutralinos.
}
\label{fig:dm}
\end{figure}

For the low $\tan\beta$ case, it is possible
to gain solutions with $A$-resonance annihilation, higgsino annihilation or stop, sbottom or stau
co-annihilation. 
For these processes to be significant, certain mass conditions needs to be fulfilled: the mass gap between $\tz_1$ and
the next ligtest sparticle needs to be within $\sim 15\%$ for coannihilation or $m_{\tz_1} \simeq 2m_A$ 
for $A$-resonance annihilation. For higgsino annihilation, $\tz_1$ needs a sizable higgsino content, which
makes it close in mass with the chargino $\tw_1$ which is a higgsino-wino mixture. 
This is illustrated in Fig.~\ref{fig:dm}, where we show solutions with $R<1.05$ 
in the {\it a}) $m_{\tst_1}\ vs.\ m_{\tz_1}$ plane, the {\it b}) $m_{\ttau_1}\ vs.\ m_{\tz_1}$ plane, 
 {\it c}) the $m_{\tb_1}\ vs.\ m_{\tz_1}$ plane,  {\it d}) the $m_{\tw_1}\ vs.\ m_{\tz_1}$ plane and 
{\it e}) the $m_A\ vs.\ m_{\tz_1}$ plane. 
In each case, the approximate needed conditions for
co-annihilation, resonance annihilation or mixed higgsino annihilation are illustrated by diagonal black lines. 
The neutralino-sbottom co-annihilation 
solutions shown in Fig.~\ref{fig:dm}{\it c}) are consistent with the results presented in Ref.~\cite{Gogoladze:2011ug}, 
where it is shown that it is not trivial to have such a scenario. 
One needs to go to the more flexible $SU(5)$ GUT scale boundary conditions to realize such solutions.

For the high $\tan\beta$ case, the value of $\Omega_{\tz_1}h^2$ tends to be very high, well over
$\Omega_{\tz_1}h^2\sim 1$. 
We see that there are no solutions with sbottom and stau coannihilation, as shown in Figs.~\ref{fig:dm}{\it b}) and {\it c}). This
is because Yukawa unification requires $A_b=A_\tau \simeq 0$ at $M_{GUT}$, which combined with large SSB mass-squared parameters
result in $\tb_1$ and $\ttau_1$ heavier than $\sim 3$~TeV, much larger than the $\tz_1$ mass. 
However, dedicated scans can find some solutions where stop co-annihilation
does occur~\cite{Gogoladze:2011be}, as shown in Fig.~\ref{fig:dm}{\it a}). 
These solutions tend to be very fine-tuned, since 
a large value of weak scale $A_t$ must be generated which pushes a normally several-TeV $\tst_1$ state
into the range where $m_{\tst_1}\sim m_{\tz_1}$.

We note here that solutions with a neutralino overabundance may be brought into accord with the
WMAP measured value of CDM by at least two methods.
\begin{enumerate} 
\item The $\tz_1$ may not in fact be the LSP,  but instead decays into a (usually) much lighter state. 
This occurs in models with mixed axion/axino ($a\ta$) dark matter~\cite{bs}, where $\tz_1\to \gamma\ta$. 
Then the neutralino abundance is converted into an axino abundance with\cite{ckkr} 
$\Omega_{\ta}h^2=\frac{m_{\ta}}{m_{\tz_1}}\Omega_{\tz_1}h^2$. If $m_{\ta}\sim$ MeV scale, then the ratio 
$m_{\ta}/m_{\tz_1}$ can reduce the putative neutralino mass abundance by many orders of magnitude.
In this case of mixed $a\ta$ CDM, axion-domination tends to be favored~\cite{bss}.
\item If additional late decaying scalar fields are present in the model, they may get produced at 
large rates via coherent oscillations. If they temporarily dominate the energy density of the universe,
and then decay to mainly SM particles, they may inject considerable entropy into the cosmic soup, 
thus diluting all relics which are present at the time of decay. 
Entropy injection can occur at large rates for instance from 
saxion production in the Peccei-Quinn augmented MSSM~\cite{shafi,bls}, or from moduli production and decay, as is 
expected in string theory~\cite{gg}. 
\end{enumerate}

In the cases where the neutralino relic abundance is too low -- 
{\it e.g.} the low $\tan\beta$ higgsino line of solutions at $\Omega_{\tz_1}h^2\sim 10^{-3}$ in Fig.~\ref{fig:Oh2} -- 
then the neutralino abundance can be augmented in the PQMSSM case where $m_{\ta}>m_{\tz_1}$, and additional
neutralinos are produced via thermal axino production and decay $\ta\to\gamma\tz_1$, or via saxion 
cascade decays~\cite{bls}. In these cases, the CDM tends to be neutralino dominated with a small component of axions.

\section{Conclusions
\label{sec:conclude}}

In this paper, we have used the Isasugra sparticle mass calculator to explore the possibility of
$b-\tau$ Yukawa-unified solutions as would be expected to occur in minimal $SU(5)$ SUSY GUT
theories. Our main assumption is that $SU(5)$ breaks at $Q=M_{GUT}$ to the MSSM as the low energy
effective theory, so that soft SUSY breaking terms are related by $SU(5)$ boundary conditions
at the GUT scale. This could occur for instance in $4-d$ models with GUT symmetry breaking via the Higgs mechanism,
or in extra dimensional GUTs where $SU(5)$ is broken via extra-dimensional compactification on
perhaps an orbifold. We search for sparticle mass spectra which maintain $f_b\simeq f_{\tau}$ at $M_{GUT}$.

We have found two sets of solutions. At low $\tan\beta$, the solutions are characterized by
$f_b\simeq f_{\tau}\sim 0.04$, while $f_t\sim 0.55$. 
These solutions have $m_{\tq}\sim m_{\tg}\sim 1-3$~TeV, and only a portion of solution space will
be accessible to LHC14 searches. 
A variety of mechanisms are available so that thermal production of neutralino CDM may occur
in the early universe and generate a dark matter abundance at the measured value.
However, these low $\tan\beta$ Yukawa-unified solutions tend to have $m_h<123$~GeV, and so may be ruled out if the
Atlas/CMS preliminary evidence for a Higgs with $m_h\simeq 125$~GeV persists.

The large $\tan\beta\sim 35-60$ solutions exhibit many of the characteristics of
$t-b-\tau$ unified solutions which are expected to occur in $SO(10)$ SUSY GUTs.
Our solutions tend to have $f_b\simeq f_{\tau}$, but with a small mis-match with $f_t$. 
These solutions tend to have multi-TeV matter scalars, but $m_{\tg}\alt 2$~TeV. 
Thus, they relax somewhat the tendency that $m_{\tg}\alt 500$~GeV as
occurs in $t-b-\tau$ unified models with gaugino mass unification and $\mu >0$.
This class of models should lead ultimately to detection of gluino pair events at 
either LHC7 or LHC14. They tend to overproduce neutralino dark matter except in the (unlikely) 
case where top-squark co-annihilation may occur. 
The cases of neutralino dark matter overproduction can be
brought into accord with astrophysical measurements by invoking further neutralino decays 
(for instance decay to an axino LSP), or via entropy dilution of any relics by saxion or moduli decays.

\section*{Acknowledgments}
This work is supported in part by the DOE Grants DE-FG02-04ER41305 (H.B.), DE-FG02-91ER40626
(I.G., S.R. and Q.S.) and DE-FG02-94ER-40823 (A.M.). This work used the Extreme Science
and Engineering Discovery Environment (XSEDE), which is supported by National Science
Foundation grant number OCI-1053575.


\begin{thebibliography}{99}
%
\bibitem{su5} H.~Georgi and S.~Glashow, \prl{32}{1974}{438};
H.~Georgi, H.~Quinn and S.~Weinberg, \prl{33}{1974}{451};
A.~Burad, J.~Ellis, M.~Gaillard and D.~Nanopoulos, \npb{135}{1978}{66}.
%
\bibitem{su5reviews}  For recent reviews,   
see R.~Mohapatra, hep-ph/9911272 (1999) and S.~Raby, in
Rept.~Prog.~Phys.~{\bf 67} (2004) 755.
%
\bibitem{witten} E.~Witten, \npb{188}{1982}{513}; R.~Kaul, \plb{109}{1982}{19}.
%
\bibitem{gauge} S.~Dimopoulos, S.~Raby and F.~Wilczek, \prd{24}{1981}{1681};
M.~Einhorn and D.R.T.~Jones, \npb{196}{1982}{475};
W.~Marciano and G.~Senjanovic, \prd{25}{1982}{3092};
U.~Amaldi, W.~de Boer and H.~Furstenau, 
\plb{260}{1991}{447};
J.~Ellis, S.~Kelley and D.~V.~Nanopoulos, \plb{260}{1991}{131};
P.~Langacker and Luo, \prd{44}{1991}{817}.
%
\bibitem{mp} H.~Murayama and A.~Pierce, \prd{65}{2002}{055009}.
%
\bibitem{pdecay}  R.~Arnowitt and P.~Nath, {\sl Phys.~Rev.~Lett.} {\bf 69}, 725
(1992); J.~Hisano, H.~Murayama and T.~Yanagida, {\sl Nucl.~Phys.~B} {\bf 402},
46 (1993)
%
\bibitem{5dguts} Y.~Kawamura, \ptp{105}{2001}{999}; 
G.~Altarelli and F.~Feruglio, \plb{511}{2001}{257}; 
L.~Hall and Y.~Nomura, \prd{64}{2001}{055003};  
A.~Hebecker and J.~March-Russell, \npb{613}{2001}{3};
A.~Kobakhidze, \plb{514}{2001}{131}.
%
\bibitem{raby_review} S.~Raby, Rept.\ Prog.\ Phys.\  {\bf 74} (2011) 036901.
%
\bibitem{pp} N.~Polonsky and A.~Pomarol, \prd{51}{1995}{6532}.
%
\bibitem{bdqt} H.~Baer, M.~Diaz, P.~Quintana and X.~Tata, 
\jhep{0004}{2000}{016}.
%
\bibitem{mtop}  The Tevatron Electroweak Working group 
(CDF and D0 Collaborations), arXiv:0803.1683.
%
\bibitem{old} B.~Ananthanarayan, G.~Lazarides and Q.~Shafi, 
\prd{44}{1991}{1613} and \plb{300}{1993}{245};
Q.~Shafi and
B.~Ananthanarayan, Trieste HEP Cosmol.1991:233-244; 
G.~Anderson {\it et al.} \prd{47}{1993}{3702} and \prd{49}{1994}{3660};
V.~Barger, M.~Berger and P.~Ohmann,   
\prd{49}{1994}{4908};
M.~Carena, M.~Olechowski, S.~Pokorski and C.~Wagner,  
Ref.~~\cite{hrs};   
B.~Ananthanarayan, Q.~Shafi and X.~Wang, \prd{50}{1994}{5980};
R.~Rattazzi and U.~Sarid, \prd{53}{1996}{1553};
T.~Blazek, M.~Carena, S.~Raby and C.~Wagner, \prd{56}{1997}{6919}; 
T.~Blazek and S.~Raby, \plb{392}{1997}{371};
T.~Blazek and S.~Raby, \prd{59}{1999}{095002};
T.~Blazek, S.~Raby and K.~Tobe, \prd{60}{1999}{113001}
and \prd{62}{2000}{055001}; S.~Profumo, \prd{68}{2003}{015006}; C.~Pallis, \npb{678}{2004}{398};
M.~Gomez, G.~Lazarides and C.~Pallis, 
\prd{61}{2000}{123512}, \npb{638}{2002}{165} and \prd{67}{2003}{097701};
U.~Chattopadhyay, A.~Corsetti and P.~Nath, \prd{66}{2002}{035003};
M.~Gomez, T.~Ibrahim, P.~Nath and S.~Skadhauge,
\prd{72}{2005}{095008}.
%
\bibitem{bf} H.~Baer and J.~Ferrandis, \prl{87}{2001}{211803}.
%
\bibitem{bdr} T.~Blazek, R.~Dermisek and S.~Raby, \prl{88}{2002}{111804};
T.~Blazek, R.~Dermisek and S.~Raby, \prd{65}{2002}{115004}.
%
\bibitem{abbbft} D.~Auto, H.~Baer, C.~Balazs, A.~Belyaev, J.~Ferrandis 
and X.~Tata, \jhep{0306}{2003}{023}.
%
\bibitem{bkss} H.~Baer, S.~Kraml, S.~Sekmen and H.~Summy, \jhep{0803}{2008}{056}.
%
\bibitem{supergut} 
  J.~Ellis, A.~Mustafayev and K.~A.~Olive,
  Eur.\ Phys.\ J.\ C {\bf 69}, 201 (2010)
  and 
  Eur.\ Phys.\ J.\ C {\bf 71}, 1689 (2011)
  [arXiv:1103.5140 [hep-ph]].
%
\bibitem{noscale} 
M. Schmaltz and W. Skiba, \prd{62}{2000}{095005} and \prd{62}{2000}{095004};
H. Baer, A. Belyaev, T. Krupovnickas and X. Tata, \prd{65}{2002}{075024};
  J.~Ellis, A.~Mustafayev and K.~A.~Olive,
  Eur.\ Phys.\ J.\ C {\bf 69}, 219 (2010)
  [arXiv:1004.5399 [hep-ph]].
%
\bibitem{imh}  J.~Feng, C.~Kolda and N.~Polonsky, \npb{546}{1999}{3}; 
J.~Bagger, J.~Feng and N.~Polonsky, \npb{563}{1999}{3};
J.~Bagger, J.~Feng, N.~Polonsky and R.~Zhang, \plb{473}{2000}{264};
H.~Baer,P.~Mercadante and X.~Tata, \plb{475}{2000}{289};
H.~Baer, C.~Balazs, M.~Brhlik, P.~Mercadante, X.~Tata and Y.~Wang,
\prd{64}{2001}{015002}.
%
\bibitem{dr3} H.~Baer, S.~Kraml and S.~Sekmen,  \jhep{0909}{2009}{005}.
%
\bibitem{mop} R.~Rattazzi and U.~Sarid, \prd {53}{1996}{1553}.~H.~Murayama,
M.~Olechowski and S.~Pokorski, Phys.~Lett.~B{\bf 371}, 57 (1996).
%
\bibitem{brs} H.~Baer, S.~Razi and Q.~Shafi, in preparation.
%
\bibitem{atlas} G.~Aad {\it et al.}  [ATLAS Collaboration],
  Phys.\ Lett.\  B {\bf 701} (2011) 398 .
%
\bibitem{lhc7} H.~Baer, V.~Barger, A.~Lessa and X.~Tata, \jhep{1006}{2010}{102} and arXiv:1112.3044.
%
\bibitem{Gogoladze:2009ug}
  I.~Gogoladze, R.~Khalid and Q.~Shafi,
  Phys.\ Rev.\  D {\bf 79}, 115004 (2009);
  I.~Gogoladze, R.~Khalid and Q.~Shafi,
  Phys.\ Rev.\  D {\bf 80}, 095016 (2009);

\bibitem{Gogoladze:2010fu}
  I.~Gogoladze, R.~Khalid, S.~Raza and Q.~Shafi,
  JHEP {\bf 1012}, 055 (2010);
  I.~Gogoladze, R.~Khalid, S.~Raza and Q.~Shafi,
  JHEP {\bf 1106} (2011) 117.
 
\bibitem{pati}
J.~C.~Pati and A.~Salam,
  Phys.\ Rev.\  D {\bf 10}, 275 (1974).

\bibitem{Profumo:2004wk}
  S.~Profumo and C.~E.~Yaguna,
  Phys.\ Rev.\  D {\bf 69}, 115009 (2004);
  D.~Feldman, Z.~Liu and P.~Nath,
  Phys.\ Rev.\  D {\bf 80}, 015007 (2009);
 H.~Baer, S.~Kraml, A.~Lessa and S.~Sekmen,
 JHEP {\bf 1002}, 055 (2010);
  D.~Feldman, Z.~Liu and P.~Nath,
  Phys.\ Rev.\  D {\bf 80}, 015007 (2009);
  M.~A.~Ajaib, T.~Li, Q.~Shafi and K.~Wang,
  JHEP {\bf 1101}, 028 (2011).

\bibitem{Bennett:2006fi}
  G.~W.~Bennett {\it et al.}  [Muon G-2 Collaboration],
  Phys.\ Rev.\  D {\bf 73}, 072003 (2006).

\bibitem{Gogoladze:2011ce}
  I.~Gogoladze, Q.~Shafi and C.~S.~Un,
  Phys.\ Lett.\ B {\bf 704}, 201 (2011)
  [arXiv:1107.1228 [hep-ph]].

\bibitem{Gogoladze:2011aa}
  I.~Gogoladze, Q.~Shafi and C.~S.~Un,
  arXiv:1112.2206 [hep-ph].

\bibitem{isajet} Isajet, by F.~Paige, S.~Protopopescu, H.~Baer and
X.~Tata, hepph/0312045; Isasugra, by 
H.~Baer, C.~H.~Chen, R.~Munroe, F.~Paige and X.~Tata, \prd{51}{1995}{1046}.
%
\bibitem{bfkp} H.~Baer, J.~Ferrandis, S.~Kraml and W.~Porod, \prd{73}{2006}{015010}.
%
\bibitem{Hisano:1992jj}
J.~Hisano, H.~Murayama, and T.~Yanagida,
  { Nucl.~Phys.} {\bf B402} (1993) 46.
Y.~Yamada,
{ Z.~Phys.} {\bf C60} (1993) 83;
 J.~L.~Chkareuli and I.~G.~Gogoladze,
  Phys.\ Rev.\  D {\bf 58}, 055011 (1998).
%
\bibitem{mv} S.~Martin and M.~Vaughn,  Phys.~Rev.~D{\bf 50}, 2282 (1994).
%
\bibitem{haber} H.~Haber and R.~Hempfling, \prd{48}{1993}{4280}.
%
\bibitem{pbmz} D.~Pierce, J.~Bagger, K.~Matchev and R.~Zhang,
Nucl.~Phys.~{\bf B491}, 3 (1997).
%
\bibitem{hrs} L.~J.~Hall, R.~Rattazzi and U.~Sarid, Phys.~Rev.~D{\bf 50}, 7048 (1994) 
%
\bibitem{kraml} G.~Belanger, S.~Kraml and A.~Pukhov, \prd{72}{2005}{015003}.
%
\bibitem{mcmc} 
  G.~Belanger, F.~Boudjema, A.~Pukhov and R.~K.~Singh,
  JHEP {\bf 0911}, 026 (2009)
  [arXiv:0906.5048 [hep-ph]].
%
\bibitem{wmap7} 
  E.~Komatsu {\it et al.}  [WMAP Collaboration],
  Astrophys.\ J.\ Suppl.\  {\bf 192}, 18 (2011)
  [arXiv:1001.4538 [astro-ph.CO]].
%
\bibitem{bmm} K.~Babu and C.~Kolda, \prl{84}{2000}{228};
A.~Dedes, H.~Dreiner and U.~Nierste, \prl{87}{2001}{251804};
J.~K.~Mizukoshi, X.~Tata and Y.~Wang, \prd{66}{2002}{115003}.
%
\bibitem{bsg} H.~Baer and M.~Brhlik, \prd{55}{1997}{4463};
H.~Baer, M.~Brhlik, D.~Castano and X.~Tata, \prd{58}{1998}{015007}.
%
\bibitem{mamoudi} D.~Eriksson, F.~Mahmoudi and O.~Stal, \jhep{0811}{2008}{035}.
%
\bibitem{bmmexp} CMS and LHCb Collaborations,   
  CMS-PAS-BPH-11-019; LHCb-CONF-2011-047; CERN-LHCb-CONF-2011-047.
%
\bibitem{hfag} D.~Asner {\it et al.}  (Heavy Flavor Averaging Group),
  arXiv:1010.1589 [hep-ex].
%
\bibitem{Gogoladze:2011be}
  I.~Gogoladze, S.~Raza and Q.~Shafi,
  Phys.\ Lett.\ B {\bf 706}, 345 (2012)
%
\bibitem{alt}W.~Altmannshofer, D.~Guadagnoli, S.~Raby and D.~M.~Straub,
  Phys.\ Lett.\  B {\bf 668} (2008) 385.
%
\bibitem{atlas_h} F.~Gianotti (ATLAS Collaboration), talk at CERN public seminar,
Dec.~13, 2011; ATLAS collaboration, ATLAS-CONF-2011-163 (2011).~
%
\bibitem{cms_h}  G.~Tonelli (CMS Collaboration), talk at CERN public seminar,
Dec.~13, 2011.
%
\bibitem{h125} H.~Baer, V.~Barger and A.~Mustafayev, arXiv:1112.3017.
%
\bibitem{gm2_th} T.~Moroi,
  \prd{53}{1996}{6565}
  [Erratum-ibid.\  D {\bf 56} (1997) 4424];
H.~Baer, C.~Balazs, J.~Ferrandis and X.~Tata,
\prd{64}{2001}{035004}.
%
\bibitem{davier} M.~Davier, A.~Hoecker, B.~Malaescu and Z.~Zhang,
  \epjc{71}{2011}{1515} 
  [arXiv:1010.4180 [hep-ph]].
%
\bibitem{lhc14}  H.~Baer, X.~Tata and J.~Woodside,
\prd{45}{1992}{142}; H.~Baer, C.~H.~Chen, F.~Paige and X.~Tata,
\prd{52}{1995}{2746} and \prd{53}{1996}{6241}; H.~Baer, C.~H.~Chen,
M.~Drees, F.~Paige and X.~Tata, \prd{59}{1999}{055014} H.~Baer,
C.~Bal\'azs, A.~Belyaev, T.~Krupovnickas and X.~Tata, 
\jhep{0306}{2003}{054}; see also, S.~Abdullin and F.~Charles,
\npb{547}{1999}{60}; S.~Abdullin {\it et al.} (CMS Collaboration),
\jphg{28}{2002}{469}; B.~Allanach, J.~Hetherington,
A.~Parker and B.~Webber, \jhep{08}{2000}{017};
H.~Baer, V.~Barger, A.~lessa and X.~Tata, \jhep{0909}{2009}{063}.
%
\bibitem{so10lhc} H.~Baer, S.~Kraml, S.~Sekmen and H.~Summy,
  JHEP {\bf 0810} (2008) 079;
H.~Baer, S.~Kraml, A.~Lessa and S.~Sekmen,
  JHEP {\bf 1002} (2010) 055.
%
\bibitem{wh} H.~Baer, V.~Barger, A.~Lessa, W.~Sreethawong and X.~Tata,
  arXiv:1201.2949 [hep-ph].
%
\bibitem{roszkowski} R.~Dermisek, S.~Raby, L.~Roszkowski and
R.~Ruiz de Austri, \jhep{0304}{2003}{037} and \jhep{0509}{2005}{029}.
%
\bibitem{codes} S.~Kraml, private communication.
%
\bibitem{isared}  H.~Baer, C.~Balazs and A.~Belyaev, \jhep{0203}{2002}{042}.
%
\bibitem{Gogoladze:2011ug} 
  I.~Gogoladze, S.~Raza and Q.~Shafi,
  arXiv:1111.6299 [hep-ph].
%
\bibitem{bs} H.~Baer and H.~Summy, \plb{666}{2008}{5};
H.~Baer, M.~Haider, S.~Kraml, S.~Sekmen and H.~Summy, JCAP{\bf 0902} (2009) 002.
%
\bibitem{ckkr}  L. Covi, J. E. Kim and L. Roszkowski, 
\prl{82}{1999}{4180}; L. Covi, H. B. Kim, J. E. Kim and 
L.Roszkowski, \jhep{0105}{2001}{033}.
%
\bibitem{bss} H.~Baer, A.~Box and H.~Summy, \jhep{0908}{2009}{080}.
%
\bibitem{shafi} G.~Lazarides, C.~Panagiotakapolous and Q.~Shafi, 
\plb{192}{1987}{323};G.~Lazarides, R.~Schaefer, D.~Seckel and Q.~Shafi, 
\npb{346}{1990}{193};J.~McDonald, \prd{43}{1991}{1063};
C.~Pallis, \app{21}{2004}{689}.
%
\bibitem{bls}  K-Y.~Choi, J.~E.~Kim, H.~M.~Lee and O.~Seto, \prd{77}{2008}{123501};
H.~Baer, S.~Kraml, A.~Lessa and S.~Sekmen,
  JCAP {\bf 1104} (2011) 039; H.~Baer and A.~Lessa, \jhep{1106}{2011}{027};
H.~Baer, A.~Lessa, S.~Rajagopalan and W.~Sreethawong, JCAP{\bf 1106} (2011) 031;
H.~Baer, A.~Lessa and W.~Sreethawong, arXiv:1110.2491.
%
\bibitem{gg}  T.~Moroi and L.~Randall, \npb{570}{2000}{455};
G.~Gelmini and P.~Gondolo, \prd{74}{2006}{023510};
G.~Gelmini, P.~Gondolo, A.~Soldatenko and C.~Yaguna, \prd{74}{2006}{083514};
G.~Gelmini, P.~Gondolo, A.~Soldatenko and C.~Yaguna, \prd{76}{2007}{015010};
B.~Acharya, K.~Bobkov, G.~Kane, P.~Kumar and J.~Shao, \prd{76}{2007}{126010}
and \prd{78}{2008}{065038};
B.~Acharya, P.~Kumar, K.~Bobkov, G.~Kane, J.~Shao and S.~Watson, \jhep{0806}{2008}{064}.
%
\end{thebibliography}
\end{document}